\documentclass[ejs]{imsart}

\RequirePackage[OT1]{fontenc}
\RequirePackage{amsthm,amsmath}
\RequirePackage[numbers]{natbib}
\RequirePackage[colorlinks,citecolor=blue,urlcolor=blue]{hyperref}
\usepackage{graphicx}
\usepackage{lineno,hyperref}
\modulolinenumbers[5]

\usepackage{amsmath,amsfonts,amssymb,amsthm}
\usepackage{graphicx,psfrag,epsf}
\usepackage{enumerate}
\usepackage{algorithm,algorithmicx,algpseudocode}
\usepackage{bm}
\usepackage{booktabs}
\usepackage{multirow} 
\usepackage{float}
\usepackage{amsfonts}
\usepackage{algcompatible,amsmath}
\usepackage{caption}
\usepackage[labelformat=simple]{subcaption}
\usepackage{afterpage}

\usepackage{dashrule}

\usepackage{amssymb}
\DeclareMathOperator{\Tr}{Tr}
\DeclareMathOperator{\Var}{Var}

\DeclareMathOperator{\Ex}{E}

\DeclareMathOperator*{\argmin}{argmin}  
\DeclareMathOperator{\diag}{diag}

\DeclareMathOperator*{\vect}{vec}

\algrenewcommand{\algorithmiccomment}[1]{\hfill$\blacktriangleright$ #1}

\pubyear{2020}
\volume{0}
\issue{0}
\firstpage{1}
\lastpage{8}
\arxiv{}

\startlocaldefs
\numberwithin{equation}{section}
\theoremstyle{plain}

\endlocaldefs

\begin{document}

\begin{frontmatter}
\title{Adaptive Smoothing Spline Estimator for the Function-on-Function Linear Regression Model}
\runtitle{AdaSS Estimator for the FoF Linear Regression Model}
\thankstext{T1}{Corresponding Author}

\begin{aug}
\author{\fnms{Fabio Centofanti} \thanksref{T1}\ead[label=e1]{fabio.centofanti@unina.it}}
\address{Department of Industrial Engineering, University of Naples Federico II, Piazzale Tecchio 80, 80125, Naples, Italy\\
	\printead{e1}}
\author{Antonio Lepore \ead[label=e2]{antonio.lepore@unina.it}}
\address{Department of Industrial Engineering, University of Naples Federico II, Piazzale Tecchio 80, 80125, Naples, Italy\\
	\printead{e2}}
\author{Alessandra Menafoglio \ead[label=e3]{alessandra.menafoglio@polimi.it}}
\address{MOX - Modelling and Scientific Computing, Department of Mathematics, Politecnico di Milano, Piazza Leonardo da Vinci 32, 20133, Milan, Italy\\
	\printead{e3}}
\author{Biagio Palumbo \ead[label=e4]{biagio.palumbo@unina.it}}
\address{Department of Industrial Engineering, University of Naples Federico II, Piazzale Tecchio 80, 80125, Naples, Italy\\
	\printead{e4}}
\author{Simone Vantini \ead[label=e5]{simone.vantini@polimi.it}}
\address{MOX - Modelling and Scientific Computing, Department of Mathematics, Politecnico di Milano, Piazza Leonardo da Vinci 32, 20133, Milan, Italy\\
	\printead{e5}}

\runauthor{Centofanti et al.}

\affiliation{Some University and Another University}

\end{aug}

\begin{abstract}
	In this paper, we propose an adaptive smoothing spline (AdaSS) estimator for the function-on-function linear regression model where each value of the response, at any domain point, depends on the full trajectory of the predictor. 
	The AdaSS estimator is obtained by the optimization of  an objective function  with two  spatially adaptive penalties, based on initial estimates of the partial derivatives of the regression coefficient function. This allows the proposed estimator to adapt more easily to the true coefficient function over regions of large curvature and not to be undersmoothed  over the remaining part of the domain.
	A novel evolutionary algorithm is developed ad hoc to obtain
	the optimization tuning parameters. 
	Extensive Monte Carlo simulations have been carried out to compare the AdaSS estimator with competitors that have  already appeared in the literature before. The results show that our proposal mostly outperforms the competitor in terms of  estimation and prediction accuracy.
	Lastly, those advantages are illustrated also on two real-data benchmark examples.

\end{abstract}


\tableofcontents
\end{frontmatter}

\section{Introduction}
Complex datasets are increasingly available due to advancements in  technology and computational power and have stimulated significant methodological developments. In this regard, functional data analysis (FDA) addresses the issue of dealing with  data that can be modeled as functions defined on a compact domain. FDA is   a thriving area of statistics and, for a comprehensive overview, the reader could refer to  \cite{ramsay2005functional,hsing2015theoretical,horvath2012inference,kokoszka2017introduction,ferraty2006nonparametric}.
In particular,  the generalization  of the classical multivariate regression analysis to the case where  the predictor and/or the response  have a functional form is referred to as functional regression and is illustrated e.g.,  in \cite{morris2015functional} and  \cite{ramsay2005functional}.
Most of the functional regression methods have been developed for models with scalar response and functional predictors (scalar-on-function regression) or  functional response and scalar predictors (function-on-scalar regression). Some  results may be found in \cite{cardot2003spline, james2002generalized,yao2010functional,muller2005generalized}.
Models where both the response and the predictor are functions, namely function-on-function (FoF) regression, have been far less studied until now.
In this work, we study  FoF linear regression models, where the response variable  function, at any
domain point, depends linearly on the full trajectory of the predictor. That is,
\begin{equation}
\label{eq_lm}
Y_i\left(t\right)=\int_{\mathcal{S}}X_{i}\left(s\right)\beta\left(s,t\right)ds+\varepsilon_{i}\left(t\right) \quad t\in\mathcal{T},
\end{equation}
for $i=1,\dots,n$. 
The pairs $\left(X_{i}, Y_i\right)$ are independent realizations of the predictor $X$ and  the response $Y$, which are assumed to be smooth random process with realizations in $L^2 (\mathcal{S})$ and $L^2 (\mathcal{T})$, i.e., the
Hilbert spaces of square integrable functions defined on the compact sets $\mathcal{S}$ and $\mathcal{T}$, respectively. Without loss of generality, the latter are also assumed  with functional mean equal to zero. 
The functions $\varepsilon_{i}$ are zero-mean random errors, independent of $X_{i}$. The function   $\beta$ is smooth  in $L^2 (\mathcal{S}\times\mathcal{T})$, i.e., the Hilbert space of bivariate square integrable functions defined on the closed intervals $\mathcal{S}\times\mathcal{T}$, and is hereinafter referred to as  \textit{coefficient function}.
For each $t\in \mathcal{T}$, the  contribution of  $X_{i}\left(\cdot\right)$ to the conditional value of  $Y_i\left(t\right)$ is  generated by   $\beta\left(\cdot,t\right)$, which works as continuous set of  weights of the predictor evaluations. 
Different methods to estimate $\beta$ in \eqref{eq_lm} have been proposed in the literature. Ramsay and Silverman \cite{ramsay2005functional} assume the estimator of $\beta$  to be in a finite dimension tensor
space spanned by two basis sets and where regularization is achieved by either truncation
or roughness penalties. (The latter  is the foundation of the method proposed in this article  as we will see below.)
Yao et al. \cite{yao2005regression}  assume the estimator of $\beta$ to be in a tensor product space generated by the   eigenfunctions of the covariance  functions of the predictor $X$ and the response $Y$, estimated by using  the principal analysis by conditional expectation (PACE) method \citep{yao2005functional}.
More recently,  Luo
and Qi \cite{luo2017function}  propose an estimation method  of the FoF linear model with multiple functional predictors based on a finite-dimensional approximation of the mean  response  obtained
by solving a penalized generalized functional eigenvalue problem.  Qi and Luo \cite{qi2018function} generalize the method in \cite{luo2017function} to the high dimensional case, where the number of covariates is much larger than the sample size (i.e., $p >> n$).
In order to  improve model flexibility and prediction accuracy,  Luo and Qi \cite{luo2019interaction} consider a FoF regression model with interaction and quadratic effects. A nonlinear FoF additive regression model with multiple functional predictors is proposed by Qi and Luo  \cite{Qi2019nonlinear}.

One of the most used estimation method is the  \textit{smoothing spline estimator} $\hat{\beta}_{SS}$ introduced by Ramsay and Silverman  \cite{ramsay2005functional}. It is obtained as the solution of the following optimization problem
\begin{multline}
\label{eq_smoothest}
\hat{\beta}_{SS}=\argmin_{\alpha \in \mathbb{S}_{k_1,k_2,M_1,M_2}}\Big\{ \sum_{i=1}^{n}\int_{\mathcal{T}}\left[Y_{i}\left(t\right)-\int_{\mathcal{S}}X_{i}\left(s\right)\alpha\left(s,t\right)ds\right]^{2}dt\\
\hspace{4cm}+\lambda_{s}\int_{\mathcal{S}}\int_{\mathcal{T}}\left(\mathcal{L}_{s}^{m_{s}}\alpha\left(s,t\right)\right)^{2}dsdt+\lambda_{t}\int_{\mathcal{S}}\int_{\mathcal{T}}\left(\mathcal{L}_{t}^{m_{t}}\alpha\left(s,t\right)\right)^{2}dsdt\Big\},
\end{multline}
where $\mathbb{S}_{k_1,k_2,M_1,M_2}$ is the tensor product space generated by the sets of B-splines of orders $k_1$ and $k_2$ associated with the  non-decreasing sequences of $M_1+2$ and  $M_2+2$ knots defined on $\mathcal{S}$ and $\mathcal{T}$, respectively. The operators $\mathcal{L}_{s}^{m_{s}}$ and $\mathcal{L}_{t}^{m_{t}}$, with $m_s\leq
k_1-1$ and $m_t\leq
k_2-1$,  are the $m_{s}$th and $m_{t}$th order linear differential operators  applied to $\alpha$ with respect to the variables $s$ and $t$, respectively. The two penalty terms on the right-hand side of \eqref{eq_smoothest} measure the roughness of the  function $\alpha$. The positive constants $\lambda_{s}$ and $\lambda_{t}$ are   generally referred to as \textit{roughness parameters} and  trade off  smoothness and goodness of  fit of the  estimator. The higher their values, the smoother the  estimator of the coefficient function.

Note that the  two penalty terms on the right-side hand of \eqref{eq_smoothest} do not depend on $s$ and $t$. Therefore, the estimator $\hat{\beta}_{SS}$  may suffer from over and under smoothing when, for instance, the true  coefficient function $ \beta $ is  wiggly or peaked only in some parts of the domain.
To solve this problem,  we consider  two adaptive roughness parameters that are allowed to vary on the domain $\mathcal{S}\times\mathcal{T}$. In this way, more flexible estimators can be obtained to improve the estimation of the   coefficient function.

Methods that  use  adaptive roughness parameters are  very popular and well established in the field of nonparametric regression,  and  are referred to as \textit{adaptive} methods. In particular, the smoothing spline estimator for nonparametric regression \citep{wahba1990spline, green1993nonparametric,eubank1999nonparametric,gu2013smoothing} has been extended by different authors to take into account the non-uniform smoothness  along the domain of the function to be estimated  \citep{ruppert2000theory,pintore2006spatially,storlie2010locally,wang2013smoothing,yang2017adaptive}.
%

In this paper,  a \textit{spatially adaptive} estimator is proposed as the solutions of the following minimization problem 
\begin{multline}
\label{eq_smoothest2}
\argmin_{\alpha \in \mathbb{S}_{k_1,k_2,M_1,M_2}}\Big\{ \sum_{i=1}^{n}\int_{\mathcal{T}}\left[Y_{i}\left(t\right)-\int_{\mathcal{S}}X_{i}\left(s\right)\alpha\left(s,t\right)ds\right]^{2}dt\\
\hspace{3cm}+\int_{\mathcal{S}}\int_{\mathcal{T}}\lambda_{s}\left(s,t\right)\left(\mathcal{L}_{s}^{m_{s}}\alpha\left(s,t\right)\right)^{2}dsdt+\int_{\mathcal{S}}\int_{\mathcal{T}}\lambda_{t}\left(s,t\right)\left(\mathcal{L}_{t}^{m_{t}}\alpha\left(s,t\right)\right)^{2}dsdt\Big\},
\end{multline}
where the  two roughness parameters $\lambda_{s}\left(s,t\right)$ and $\lambda_{t}\left(s,t\right)$ are functions that produce different amount of penalty, and, thus, allow the estimator to spatially adapt, i.e.,   to accommodate varying degrees of roughness  over the domain $\mathcal{S}\times\mathcal{T}$. Therefore, the model may  accommodate the local behavior of $\beta$ by imposing a heavier penalty in regions of lower smoothness.
Because  $\lambda_{s}\left(s,t\right)$ and $\lambda_{t}\left(s,t\right)$ are intrinsically infinite dimensional, their specification could be rather complicated without further assumptions.

The proposed  estimator is applied to FoF linear regression model reported in \eqref{eq_lm}, and is referred to as adaptive smoothing spline (AdaSS) estimator. It is obtained as  the solution of the optimization problem in \eqref{eq_smoothest2}, with $\lambda_{s}\left(s,t\right)$ and $\lambda_{t}\left(s,t\right)$  chosen based on an initial estimate of the partial derivatives $\mathcal{L}_{s}^{m_{s}}\alpha\left(s,t\right)$ and $\mathcal{L}_{t}^{m_{t}}\alpha\left(s,t\right)$.
The rationale behind this choice is to allow the  contribution of $\lambda_{s}\left(s,t\right)$ and $\lambda_{t}\left(s,t\right)$,   to  the penalties in \eqref{eq_smoothest2},  to be small over regions where the initial estimate has large $m_s$th and $m_t$th curvatures (i.e.,  partial derivatives), respectively.
This can be regarded as an extension to the FoF linear regression model of the idea of  Storlie
et al. \cite{storlie2010locally}  and Abramovich and Steinberg  \cite{abramovich1996improved}.
Moreover, to overcome some limitations of the most famous  grid-search method \citep{bergstra2011algorithms}, a new evolutionary algorithm is proposed for the choice of the unknown parameters, needed to compute the AdaSS estimator.

The rest of the paper is organized as follows. In Section \ref{sec_ASS}, the proposed estimator is presented. Computational issues involved in the AdaSS estimator calculation are discussed in Section \ref{sec_derivation} and Section \ref{sec_tuning}. In Section \ref{sec_simulation}, by means of a Monte Carlo simulation study, the performance of the proposed estimator are compared  with those achieved by competing estimators  already appeared in the literature.
Lastly, two real-data examples are presented in Section \ref{sec_real} to illustrate the practical applicability of the proposed estimator.
The conclusion is in Section \ref{sec_conclusion}.

\section{The Adaptive Smoothing Spline Estimator}

\subsection{The Estimator}
\label{sec_ASS}
The AdaSS estimator $\hat{\beta}_{AdaSS}$ is defined as the solution  of the optimization problem in \eqref{eq_smoothest2} where the two roughness parameters $\lambda_{s}\left(s,t\right)$ and $\lambda_{t}\left(s,t\right)$ are as follows 
\begin{align*}
\lambda_{s}\left(s,t\right)=\lambda^{AdaSS}_{s}\frac{1}{\left(|\widehat{\beta_{s}^{m_{s}}}\left(s,t\right)|+\delta_s\right)^{\gamma_s}}\\
\lambda_{t}\left(s,t\right)=\lambda^{AdaSS}_{t}\frac{1}{\left(|\widehat{\beta_{t}^{m_{t}}}\left(s,t\right)|+\delta_t\right)^{\gamma_t}}
\end{align*}
that is,
\begin{multline}
\label{eq_ASS}
\hspace{-3cm}\hat{\beta}_{AdaSS}=\argmin_{\alpha \in \mathbb{S}_{k_1,k_2,M_1,M_2}}\Big\{ \sum_{i=1}^{n}\int_{\mathcal{T}}\left[Y_{i}\left(t\right)-\int_{\mathcal{S}}X_{i}\left(s\right)\alpha\left(s,t\right)ds\right]^{2}dt\\
+\lambda^{AdaSS}_{s}\int_{\mathcal{S}}\int_{\mathcal{T}}\frac{1}{\left(|\widehat{\beta_{s}^{m_{s}}}\left(s,t\right)|+\delta_s\right)^{\gamma_s}}\left(\mathcal{L}_{s}^{m_{s}}\alpha\left(s,t\right)\right)^{2}dsdt\\
+\lambda^{AdaSS}_{t}\int_{\mathcal{S}}\int_{\mathcal{T}}\frac{1}{\left(|\widehat{\beta_{t}^{m_{t}}}\left(s,t\right)|+\delta_t\right)^{\gamma_t}}\left(\mathcal{L}_{t}^{m_{t}}\alpha\left(s,t\right)\right)^{2}dsdt\Big\},
\end{multline}
for some tuning parameters $\lambda^{AdaSS}_{s},\delta_s,\gamma_s,\lambda^{AdaSS}_{t},\delta_t,\gamma_t\geq 0$ and  $\widehat{\beta_{s}^{m_{s}}}$ and $\widehat{\beta_{t}^{m_{t}}}$  initial estimates of $\mathcal{L}_{s}^{m_{s}}\beta$ and $\mathcal{L}_{t}^{m_{t}}\beta$, respectively.
Note that the two roughness parameters $\lambda_{s}$ and $\lambda_{t}$ assume large values over domain regions where $\widehat{\beta_{s}^{m_{s}}}$ and $\widehat{\beta_{t}^{m_{t}}}$ are small.
Therefore,   in the  right-hand side  of \eqref{eq_ASS}, $ (\mathcal{L}_{s}^{m_{s}}\alpha)^2 $ and $( \mathcal{L}_{t}^{m_{t}}\alpha)^2 $ are  weighted through the inverse of $|\widehat{\beta_{s}^{m_{s}}}|$ and $|\widehat{\beta_{t}^{m_{t}}}|$.
That is, over domain regions where $\widehat{\beta_{s}^{m_{s}}}$ and $\widehat{\beta_{t}^{m_{t}}}$ are small, $ (\mathcal{L}_{s}^{m_{s}}\alpha)^2 $ and $( \mathcal{L}_{t}^{m_{t}}\alpha)^2 $ have larger weights  than over those regions where $\widehat{\beta_{s}^{m_{s}}}$ and $\widehat{\beta_{t}^{m_{t}}}$ are large.
For this reasons,  the  final estimator is able to  adapt to the coefficient function over  regions of large curvature without  over smoothing it over regions where the  $m_s$th and $ m_t $th curvatures  are small.

The constants $\delta_s$ and $\delta_t$ allow $\hat{\beta}_{AdaSS}$  not to have $m_s$th and $m_t$th-order  inflection points  at the same location of $\widehat{\beta_{s}^{m_{s}}}$ and $\widehat{\beta_{t}^{m_{t}}}$, respectively. Indeed, when $\delta_s$ and $\delta_t$ are set to zero, where $\widehat{\beta_{s}^{m_{s}}}=0$ and $\widehat{\beta_{t}^{m_{t}}}=0$ ($m_s$th and $m_t$th-order  inflection points), the corresponding penalties go to infinite, and, thus, $\mathcal{L}_{s}^{m_{s}}\alpha\left(s,t\right)$ and $\mathcal{L}_{t}^{m_{t}}\alpha\left(s,t\right)$ become zero in accordance with the minimization problem.
Therefore, the presence of $\delta_s$ and $\delta_t$   makes $\hat{\beta}_{AdaSS}$ more robust against the choice of the initial estimate of the  linear differential operators applied to $\beta$ with respect to $s$ and $t$. Finally, $\gamma_s$ and $\gamma_t$ control the amount of weight placed in $\widehat{\beta_{s}^{m_{s}}}$ and $\widehat{\beta_{t}^{m_{t}}}$, whereas  $\lambda^{AdaSS}_{s}$ and $\lambda^{AdaSS}_{t}$ are smoothing parameters.
The solution of the optimization problem in \eqref{eq_ASS} can be obtained in closed form if  the penalty terms are approximated as  described in Section \ref{sec_derivation}.
There are several  choices for the  initial estimates $\widehat{\beta_{s}^{m_{s}}}$ and $\widehat{\beta_{t}^{m_{t}}}$. As  in   \cite{abramovich1996improved}, we suggest to apply the $m_s$th and $m_t$th order linear differential operator to the  smoothing spline estimator $\hat{\beta}_{SS}$ in \eqref{eq_smoothest}.

\subsection{The Derivation of the AdaSS Estimator}
\label{sec_derivation}

The minimization in \eqref{sec_ASS} is carried out over $\alpha\in\mathbb{S}_{k_1,k_2,M_1,M_2}$. This implicitly means that we are approximating $\beta$ as follows
\begin{equation}
\label{eq_betaapp0}
\beta\left(s,t\right)\approx\tilde{\beta}\left(s,t\right)=\sum_{i=1}^{M_1+k_1}\sum_{j=1}^{M_2+k_2}b_{ij}\psi^{s}_{i}\left(s\right)\psi^{t}_{j}\left(t\right)=\bm{\psi}^{s}\left(s\right)^{T}\bm{B}\bm{\psi}^{t}\left(t\right) \quad s\in \mathcal{S}, t\in\mathcal{T},
\end{equation}
where  $\bm{B}=\lbrace b_{ij}\rbrace\in\mathbb{R}^{M_1+k_1\times M_2+k_2}$. The two sets  $\bm{\psi}^{s}=\left(\psi^{s}_{1},\dots,\psi^{s}_{M_1+k_1}\right)^{T}$ and $\bm{\psi}^{t}=\left(\psi^{t}_{1},\dots,\psi^{t}_{M_2+k_2}\right)^{T}$ are B-spline functions of order $k_1$ and $k_2$ and non-decreasing knots sequences $\Delta^{s}=\lbrace s_{0},s_{1},\dots,s_{M_1},s_{M_1+1}\rbrace$ and $\Delta^{t}=\lbrace t_{0},t_{1},\dots,t_{M_2},t_{M_2+1}\rbrace$, defined on $\mathcal{S}=\left[s_0,s_{M_1+1}\right]$ and $\mathcal{T}=\left[t_0,t_{M_2+1}\right]$, respectively, that generate $\mathbb{S}_{k_1,k_2,M_1,M_2}$.
Thus, estimating $\beta$ in \eqref{sec_ASS} means estimating $\bm{B}$. 
Let $\alpha\left(s,t\right)=\bm{\psi}^{s}\left(s\right)^{T}\bm{B}_{\alpha}\bm{\psi}^{t}\left(t\right)$, $s\in \mathcal{S}, t\in\mathcal{T}$, in $\mathbb{S}_{k_1,k_2,M_1,M_2}$, where  $\bm{B}_{\alpha}=\lbrace b_{\alpha,ij}\rbrace\in\mathbb{R}^{M_1+k_1\times M_2+k_2}$. Then,   the first term of the right-hand side of \eqref{eq_ASS} may be rewritten  as (see \cite{ramsay2005functional}, pag 291-293, for the derivation)
\begin{equation}
\label{eq_sse}
\sum_{i=1}^{n}\int_{\mathcal{T}}\left[Y_{i}\left(t\right)-\int_{\mathcal{S}}X_{i}\left(s\right)\alpha\left(s,t\right)ds\right]^{2}dt=\sum_{i=1}^{n}\int_{\mathcal{T}}Y_{i}\left(t\right)^{2}dt-2\Tr\left[\bm{X}\bm{B}_{\alpha}\bm{Y}^{T}\right]+\Tr\left[\bm{X}^{T}\bm{X}\bm{B}_{\alpha}\bm{W}_{t}\bm{B}_{\alpha}^{T}\right],
\end{equation}
where $\bm{X}=\left(\bm{X}_{1},\dots,\bm{X}_{n}\right)^{T}$, with $\bm{X}_{i}=\int_{\mathcal{S}}X_{i}\left(s\right)\bm{\psi}^{s}\left(s\right)ds$, $\bm{Y}=\left(\bm{Y}_{1},\dots,\bm{Y}_{n}\right)^{T}$ with $\bm{Y}_{i}=\int_{\mathcal{T}}Y_{i}\left(t\right)\bm{\psi}^{t}\left(t\right)dt$, and  $\bm{W}_{t}=\int_{\mathcal{T}}\bm{\psi}^{t}\left(t\right)\bm{\psi}^{t}\left(t\right)^{T}dt$.
The term $\Tr\left[\bm{A}\right]$ denotes the trace of a square matrix $\bm{A}$.

In order to simplify the integrals in the two penalty terms on the right-hand side of \eqref{eq_ASS}, and thus obtain a linear form in $\bm{B}_{\alpha}$, we consider, for $ s\in \mathcal{S}$ and $t\in\mathcal{T}$, the following approximations of  $\widehat{\beta_{s}^{m_{s}}}$ and $\widehat{\beta_{t}^{m_{t}}}$  
\begin{equation}
\label{eq_app1}
\widehat{\beta_{s}^{m_{s}}}\left(s,t\right)\approx\sum_{i=0}^{L_s}\sum_{j=0}^{L_t}\widehat{\beta_{s}^{m_{s}}}\left(\tau_{s,i+1},\tau_{t,j+1}\right)I_{\left[\left(\tau_{s,i},\tau_{s,i+1}\right)\times \left(\tau_{t,j},\tau_{t,j+1}\right)\right]}\left(s,t\right),
\end{equation}
and
\begin{equation}
\label{eq_app2}
\widehat{\beta_{t}^{m_{t}}}\left(s,t\right)\approx\sum_{i=0}^{L_s}\sum_{j=0}^{L_t}\widehat{\beta_{t}^{m_{t}}}\left(\tau_{s,i+1},\tau_{t,j+1}\right)I_{\left[\left(\tau_{s,i},\tau_{s,i+1}\right)\times \left(\tau_{t,j},\tau_{t,j+1}\right)\right]}\left(s,t\right),
\end{equation}
where $\Theta^{s}=\lbrace \tau_{s,0},\tau_{s,1},\dots\tau_{s,L_s},\tau_{s,L_s+1}\rbrace$ and $\Theta^{t}=\lbrace \tau_{t,0},\tau_{t,1},\dots\tau_{t,L_t},\tau_{t,L_t+1}\rbrace$ are non increasing knot sequences with $\tau_{s,0}=s_0$, $\tau_{s,L_s+1}=s_{M_1+1}$, $\tau_{t,0}=t_0$, $\tau_{t,L_t+1}=t_{M_2+1}$, and $I_{\left[a\times b\right]}\left(z_1,z_2\right)=1$ for $\left(z_1,z_2\right)\in\left[a\times b\right] $ and zero elsewhere.
In \eqref{eq_app1} and \eqref{eq_app2}, we are assuming that $\widehat{\beta_{s}^{m_{s}}}$ and $\widehat{\beta_{t}^{m_{t}}}$ are well approximated by a piecewise constant function, whose values are constant on rectangles defined by the two knot sequences $\Theta^{s}$ and $\Theta^{t}$.
It can be easily proved, by following  Schumaker \cite{schumaker2007spline} (pag. 491, Theorem 12.7), that the approximation error in both cases goes to zero as the mesh widths $\overline{\delta}^{s}=\max_{i}\left(\tau_{s,i+1}-\tau_{s,i}\right)$ and  $\overline{\delta}^{t}=\max_{j}\left(\tau_{t,j+1}-\tau_{t,g}\right)$ go to zero. Therefore, $\widehat{\beta_{s}^{m_{s}}}$ and $\widehat{\beta_{t}^{m_{t}}}$ can be exactly recovered by uniformly increasing the number of knots $L_s$ and $L_t$.
In this way, the two penalties on the right-hand side of \eqref{eq_ASS} can be rewritten as (\ref{sec_app1})
\begin{multline}
\label{eq_pen1}
\lambda^{AdaSS}_{s}\int_{\mathcal{S}}\int_{\mathcal{T}}\frac{1}{\left(|\widehat{\beta_{s}^{m_{s}}}\left(s,t\right)|+\delta_s\right)^{\gamma_s}}\left(\mathcal{L}_{s}^{m_{s}}\alpha\left(s,t\right)\right)^{2}dsdt\\
\hspace{2cm}\approx \lambda^{AdaSS}_{s}\sum_{i=1}^{L_s+1}\sum_{j=1}^{L_t+1}d^{s}_{ij}\Tr\left[ \bm{B}_{\alpha}^{T}\bm{R}_{s,i}\bm{B}_{\alpha}\bm{W}_{t,j}\right]
\end{multline}
and
\begin{multline}
\label{eq_pen2}
\lambda^{AdaSS}_{t}\int_{\mathcal{S}}\int_{\mathcal{T}}\frac{1}{\left(|\widehat{\beta_{s}^{m_{s}}}\left(s,t\right)|+\delta_t\right)^{\gamma_t}}\left(\mathcal{L}_{t}^{m_{t}}\alpha\left(s,t\right)\right)^{2}dsdt\\
\hspace{2cm}\approx \lambda^{AdaSS}_{t}\sum_{i=1}^{L_s+1}\sum_{j=1}^{L_t+1}d^{t}_{ij}\Tr\left[ \bm{B}_{\alpha}^{T}\bm{W}_{s,i}\bm{B}_{\alpha}\bm{R}_{t,j}\right],
\end{multline}
where  $\bm{W}_{s,i}=\int_{\left[\tau_{s,i-1},\tau_{s,i}\right]}\bm{\psi}^{s}\left(s\right)\bm{\psi}^{s}\left(s\right)^{T}ds$,  $\bm{W}_{t,j}=\int_{\left[\tau_{t,j-1},\tau_{t,j}\right]}\bm{\psi}^{t}\left(t\right)\bm{\psi}^{t}\left(t\right)^{T}dt$, $\bm{R}_{s,i}=\int_{\left[\tau_{s,i-1},\tau_{s,i}\right]}\mathcal{L}_{s}^{m_{s}}\left[\bm{\psi}^{s}\left(s\right)\right]\mathcal{L}_{s}^{m_{s}}\left[\bm{\psi}^{s}\left(s\right)\right]^{T}ds$ and $\bm{R}_{t,j}=\int_{\left[\tau_{t,j-1},\tau_{t,j}\right]}\mathcal{L}_{t}^{m_{t}}\left[\bm{\psi}^{t}\left(t\right)\right]\mathcal{L}_{t}^{m_{t}}\left[\bm{\psi}^{t}\left(t\right)\right]^{T}dt$, and $d^{s}_{ij}=\Big\{\frac{1}{\left(|\widehat{\beta_{s}^{m_{s}}}\left(\tau_{s,i},\tau_{t,j}\right)|+\delta_s\right)^{\gamma_s}}\Big\}$ and $d^{t}_{ij}=\Big\{\frac{1}{\left(|\widehat{\beta_{t}^{m_{t}}}\left(\tau_{s,i},\tau_{t,j}\right)|+\delta_t\right)^{\gamma_t}}\Big\}$, for $i=1,\dots,L_s+1$ and $j=1,\dots,L_t+1$.

The  optimization problem in \eqref{eq_ASS} can be then approximated with the following
\begin{align}
\label{eq_slassoobj}
\hspace{-1cm}\hat{\bm{B}}_{AS}\approx\argmin_{\bm{B}_{\alpha} \in \mathbb{R}^{\left(M_1+k_1\right)\times \left(M_2+k_2\right)}}\Big\{&\sum_{i=1}^{n}\int_{\mathcal{T}}Y_{i}\left(t\right)^{2}dt-2\Tr\left[\bm{X}\bm{B}_{\alpha}\bm{Y}^{T}\right]+\Tr\left[\bm{X}^{T}\bm{X}\bm{B}_{\alpha}\bm{W}_{t}\bm{B}_{\alpha}^{T}\right]\nonumber\\&\hspace{-1cm}+\sum_{i=1}^{L_s+1}\sum_{j=1}^{L_t+1}\left(\lambda^{AdaSS}_{s}d^{s}_{ij}\Tr\left[ \bm{B}_{\alpha}^{T}\bm{R}_{s,i}\bm{B}_{\alpha}\bm{W}_{t,j}\right]+\lambda^{AdaSS}_{t}d^{t}_{ij}\Tr\left[ \bm{B}_{\alpha}^{T}\bm{W}_{s,i}\bm{B}_{\alpha}\bm{R}_{t,j}\right]\right)\Big\},
\end{align}
or by vectorization as
\begin{align}
\label{eq_adsmovec}
\hat{\bm{b}}_{AS}\approx\argmin_{\bm{b}_{\alpha} \in \mathbb{R}^{\left(M_1+k_1\right) \left(M_2+k_2\right)}}\Big\{&-2\vect\left(\bm{X}^{T}\bm{Y}\right)^{T}\bm{b}_{\alpha}+\bm{b}_{\alpha}^{T}\left(\bm{W}_{t}\otimes\bm{X}^{T}\bm{X}\right)\bm{b}_{\alpha}\nonumber\\
&\sum_{i=1}^{L_s+1}\sum_{j=1}^{L_t+1}\left(\lambda^{AdaSS}_{s}d^{s}_{ij}\bm{b}_{\alpha}^{T}\bm{L}_{wr,ij}\bm{b}_{\alpha}+\lambda^{AdaSS}_{t}d^{t}_{ij}\bm{b}_{\alpha}^{T}\bm{L}_{rw,ij}\bm{b}_{\alpha}\right)\bigg\},
\end{align}
where $\hat{\bm{b}}_{AS}=\vect\left(\hat{\bm{B}}_{AS}\right)$, $\bm{L}_{rw,ij}=\left(\bm{R}_{t,j}\otimes \bm{W}_{s,i}\right)$ and   $\bm{L}_{wr,ij}=\left(\bm{W}_{t,j}\otimes \bm{R}_{s,i}\right)$, for $i=1,\dots,L_s+1$ and $j=1,\dots,L_t+1$. For  a matrix $\bm{A}\in \mathbb{R}^{j\times k}$, $\vect(\bm{A})$ indicates the vector of length $jk$ obtained by writing the matrix $\bm{A}$ as a vector column-wise, and $\otimes$ is the Kronecker product.
Because the matrices $\bm{W}_{t}$, $\bm{L}_{wr,ij}$ and $\bm{L}_{rw,ij}$  for $i=1,\dots,L_s+1$ and $j=1,\dots,L_t+1$ are positive definite and by assuming that $\bm{X}^{T}\bm{X}$ is positive definite, then the minimizer of the optimization problem in \eqref{eq_adsmovec} exists, is unique  and has the following expression \citep{boyd2004convex}
\begin{align}
\label{eq_adsmosol}
\hat{\bm{b}}_{AdaSS}\approx \left[\left(\bm{W}_{t}\otimes\bm{X}^{T}\bm{X}\right)+\sum_{i=1}^{L_s+1}\sum_{j=1}^{L_t+1}\left(\lambda^{AdaSS}_{s}d^{s}_{ij}\bm{L}_{wr,ij}+\lambda^{AdaSS}_{t}d^{t}_{ij}\bm{L}_{rw,ij}\right)\right]^{-1}\vect\left(\bm{X}^{T}\bm{Y}\right).
\end{align}
To obtain $\hat{\bm{b}}_{AdaSS}$ in \eqref{eq_adsmosol} the tuning parameters $\lambda^{AdaSS}_{s},\delta_s,\gamma_s,\lambda^{AdaSS}_{t},\delta_t,\gamma_t$ must be opportunely  chosen. This issue is discussed in Section \ref{sec_tuning}.

\subsection{The Algorithm for the Parameter Selection}
\label{sec_tuning}
There are some tuning parameters in the optimization problem  \eqref{eq_adsmovec} that must be chosen to obtain the AdaSS estimator.
Usually, the tensor product space $\mathbb{S}_{k_1,k_2,M_1,M_2}$ is chosen with $k_1=k_2=4$, i.e., cubic B-splines, and equally spaced knot sequences. Although the choice of $M_1$ and $M_2$ is not crucial \citep{cardot2003spline}, it should allow the final estimator to capture the local behaviour of the coefficient function $\beta$, that is, $M_1$ and $M_2$ should be sufficiently large.
The smoothness of the final estimator is controlled by the two penalty terms on the right-hand side of \eqref{eq_adsmovec}.

The tuning parameters $\lambda^{AdaSS}_{s},\delta_s,\gamma_s,\lambda^{AdaSS}_{t},\delta_t,\gamma_t$ could be fixed by using the conventional $K$-fold cross validation (CV) \citep{trevor2009elements}, where the  combination of parameters to be explored is chosen by means of the classic grid search method \citep{trevor2009elements}. That is 
an exhaustive searching through a manually specified subset of the tuning parameter space \citep{bergstra2012random}.
Although, in our setting, grid search is embarrassingly parallel \citep{herlihy2011art}, it is not scalable because it suffers from the curse of dimensionality. However, even if this is beyond the scope of the present work, note that the number of combinations to explore grows exponentially with the number of tuning parameters and makes unsuitable  the application of the proposed method   to the FoF linear model in the case of multiple predictors. 
Then, to facilitate the use of the proposed method by practitioners, in what follows, we proposed a novel evolutionary algorithm for tuning parameter selection, referred to as \textit{evolutionary algorithm for adaptive smoothing estimator} (EAASS) inspired by the \textit{population based training} (PBT)  introduced by Jaderberg et al.  \cite{jaderberg2017population}.
The  PBT algorithm was introduced to address the issue of hyperparameter optimization for neural networks. It bridges and extends parallel search method (e.g., grid search and random search) with sequential optimization method (e.g., hand tuning and Bayesian optimization).
The former runs many parallel optimization processes, for different combinations of hyperparameter values, and, then chooses the combination that shows the best performance.
The latter performs several steps of few parallel optimizations, where, at each step, information coming from the previous step is used   to identify the combinations of hyperparameter values to explore.
For further details on the PBT algorithm the readers should refer to \cite{jaderberg2017population}, where the authors demonstrated its  effectiveness and wide applicability.
The pseudo code of the EAASS algorithm is given  in  Algorithm \ref{al_1}.
\begin{algorithm}[H]
	\caption{EAASS algorithm}  \label{al_1}
	\begin{algorithmic}[1]
		\STATE{Choose the initial population $\mathcal{P}=\lbrace p_i\rbrace$ of combinations of tuning parameter values}
		\STATE{Obtain the set $\mathcal{V}=\lbrace v_i\rbrace$ of estimated prediction errors corresponding to $\mathcal{P}$}
		\REPEAT
		\STATE{Identify the set $\mathcal{Q}\subseteq \mathcal{P}$ and the corresponding $\mathcal{Z}\subseteq \mathcal{V}$}\Comment{\textit{exploitation} }
		\FOR{$p_i\in \mathcal{Q}$}\Comment{\textit{exploration}}
		\STATE{Obtain the new combination of tuning parameter values, $p'_i$} 
		\STATE{Obtain the new estimated prediction error  $v'_i$  corresponding to $p'_i$}
		\ENDFOR
		\STATE{Define $\mathcal{Q}'=\lbrace p'_i\rbrace$ and $\mathcal{Z}'=\lbrace v'_i\rbrace$}
		\STATE {Set $\mathcal{P}=\mathcal{P}\setminus \mathcal{Q} \cup \mathcal{Q}'$  and $\mathcal{V}=\mathcal{V}\setminus \mathcal{Z} \cup \mathcal{Z}'$}
		\UNTIL{The stopping condition is met}
		\STATE {Return $p_i\in \mathcal{P}$ with the highest $v_i\in\mathcal{V}$} 
	\end{algorithmic}
\end{algorithm}

The first step is the identification of an initial population $\mathcal{P}$ of tuning parameter combinations $p_i$s. This can be done, for each combination and each tuning parameter, by randomly selecting a value in a pre-specified range.
Then, the set $\mathcal{V}$ of  estimated prediction errors $v_i$s corresponding to $\mathcal{P}$ is obtained by means of $K$-fold CV. We choose a subset $\mathcal{Q}$ of  $\mathcal{P}$, by following a given  exploitation strategy and, thus, the corresponding subset $\mathcal{Z}$ of $\mathcal{V}$. 
A typical exploitation strategy is the \textit{truncation selection}, where the worse $r\%$, for $0\leq r \leq 100$, of $\mathcal{P}$, in terms of estimated prediction error,  is substituted by elements randomly sampled from the remaining $(100-r)\%$ part of the current population \citep{jaderberg2017population}. 
Then the following step consists of an  exploration strategy where the tuning parameter combinations in $\mathcal{Q}$ are substituted by new ones. The simulation study in Section \ref{sec_simulation} and  the real-data Examples in Section \ref{sec_real} are based on a \textit{perturbation} where each tuning parameter value of the given combination is randomly perturbed by a factor of 1.2 or 0.8.
The exploitation and exploration phases are repeated until a stopping condition is met, e.g, maximum number of iterations. Other exploration and exploitation strategies can be found in \cite{back1997handbook}. 
At last, the selected tuning parameter combination is obtained as an element of $\mathcal{P}$ that achieves  the lowest  estimated prediction error.
As a remark,  in our trials  the AdaSS estimator works quite well with $\delta_s=\delta_s^*\max|\widehat{\beta_{s}^{m_{s}}}\left(s,t\right)|$ and $\delta_t=\delta_t^*\max|\widehat{\beta_{t}^{m_{t}}}\left(s,t\right)|$, for $0\leq \delta_s^*,\delta_t^*\leq 0.1$.
\section{Simulation Study}
\label{sec_simulation}
In this section, the performance of the AdaSS estimator is assessed  on several simulated datasets.
In particular, we compare the  AdaSS estimator with cubic B-splines and $m_s=m_t=2$ with five competing methods that represent  the state of the art in the FoF liner regression model estimation.
The first two are those proposed by Ramsay and
Silverman \cite{ramsay2005functional}. The first one, hereinafter referred to as SMOOTH estimator, is the smoothing spline estimator  described in \eqref{eq_smoothest}, whereas, the second one, hereinafter referred to as TRU estimator,  assumes that the  coefficient function
is in a finite dimensional tensor product  space generate by two sets of B-splines with regularization achieved by choosing the space dimension.
Then, we consider also the estimator proposed by Yao et al.  \cite{yao2005regression} and Canale and Vantini  \cite{canale2016constrained}. The former is based on the functional principal component decomposition, and is  hereinafter referred to as PCA estimator, while the latter relies on a ridge type penalization, hereinafter referred to as RIDGE estimator. Lastly, as the fifth alternative, we explore the estimator proposed by Luo and Qi  \cite{luo2017function}, hereinafter referred to as SIGCOMP. 
Moreover, the AdaSS estimator with cubic B-splines and $m_s=m_t=2$ is considered.
For illustrative purposes, we also consider a version of the AdaSS estimator, referred to AdaSStrue, whose roughness parameters are calculated by assuming that the true coefficient function is known. Obviously, the AdaSStrue has not a practical meaning because the true coefficient function is never known. However, it allows one to better understand the influence of the initial estimates of the partial derivatives on the AdaSS performance.
All the unknown parameters of the competing methods considered are chosen by means of $10$-fold CV. 
The tuning parameters of the  AdaSS and AdaSStrue estimators are chosen through the EAASS algorithm. The set $\mathcal{P}$ is obtained by using $10$-fold CV, the exploitation and exploration phases are as described in Section \ref{sec_tuning} and a   maximum number of iterations equal to 15 is set as stopping condition.
For each simulation, a training sample of $n$ observations is generated along with a test set $T$ of size $N=4000$. They are used to estimate $\beta$ and to test the predictive performance of the estimated model, respectively. Three different sample sizes are considered, viz., $n=100,500,1000$.
The estimation accuracy of the estimators are assessed by using the \textit{integrated squared error } (ISE) defined as 
\begin{equation}
\text{ISE}=\frac{1}{A}\int_{\mathcal{S}}\int_{\mathcal{T}}\left(\hat{\beta}\left(s,t\right)-\beta\left(s,t\right)\right)^{2}dsdt ,
\end{equation}
where $A$ is the measure of $\mathcal{S}\times\mathcal{T}$. The ISE aims to measure the estimation error of $\hat{\beta}$ with respect to $\beta$. Whereas, the predictive accuracy  is measured through the \textit{prediction mean squared error} (PMSE) defined as 
\begin{equation}
\text{PMSE}=\frac{1}{N}\sum_{\left(X,Y\right)\in T} \int_{\mathcal{T}}\left(Y\left(t\right)-\int_{\mathcal{S}}X\left(s\right)\hat{\beta}\left(s,t\right)ds\right)^{2}dt.
\end{equation}
The observations in the test
set are centred by subtracting to each observation the corresponding  sample mean function estimated in the training set.
The observations in the training and test sets are obtained as follows. The covariate $X_i$ and the errors $\varepsilon_i$ are generated as linear combination of  cubic B-splines, $\Psi_{i}^{x}$ and $\Psi_{i}^{\varepsilon}$,  with evenly spaced knots, i.e., $X_{i}=\sum_{j=1}^{32}x_{ij}\Psi_{i}^{x}$ and $\varepsilon_{i}=k\sum_{j=1}^{20}e_{ij}\Psi_{i}^{\varepsilon}$.
The coefficients $x_{ij}$ and $e_{ij}$, for $i=1,\dots,n$, $j=1,\dots,32$ and $j=1,\dots,20$, are independent realizations of standard normal random variable and the  numbers of basis have been randomly chosen between 10 and 50. The constant $k$ is chosen such that the  signal-to-noise ratio  $SN\doteq\int_{\mathcal{T}}\Var_{X}[\Ex\left( Y_i|X_i\right)]/\int_{\mathcal{T}}\Var\left(\varepsilon_i\right)$ is  equal to 4, where $\Var_{X}$ is the variance with respect to the random covariate $X$. Then, given the coefficient function $\beta$, the response $Y_i$ is obtained.
\subsection{Mexican Hat Function}
\label{sec_mex}
\begin{figure}
	
	\centering
	\includegraphics[width=1\textwidth]{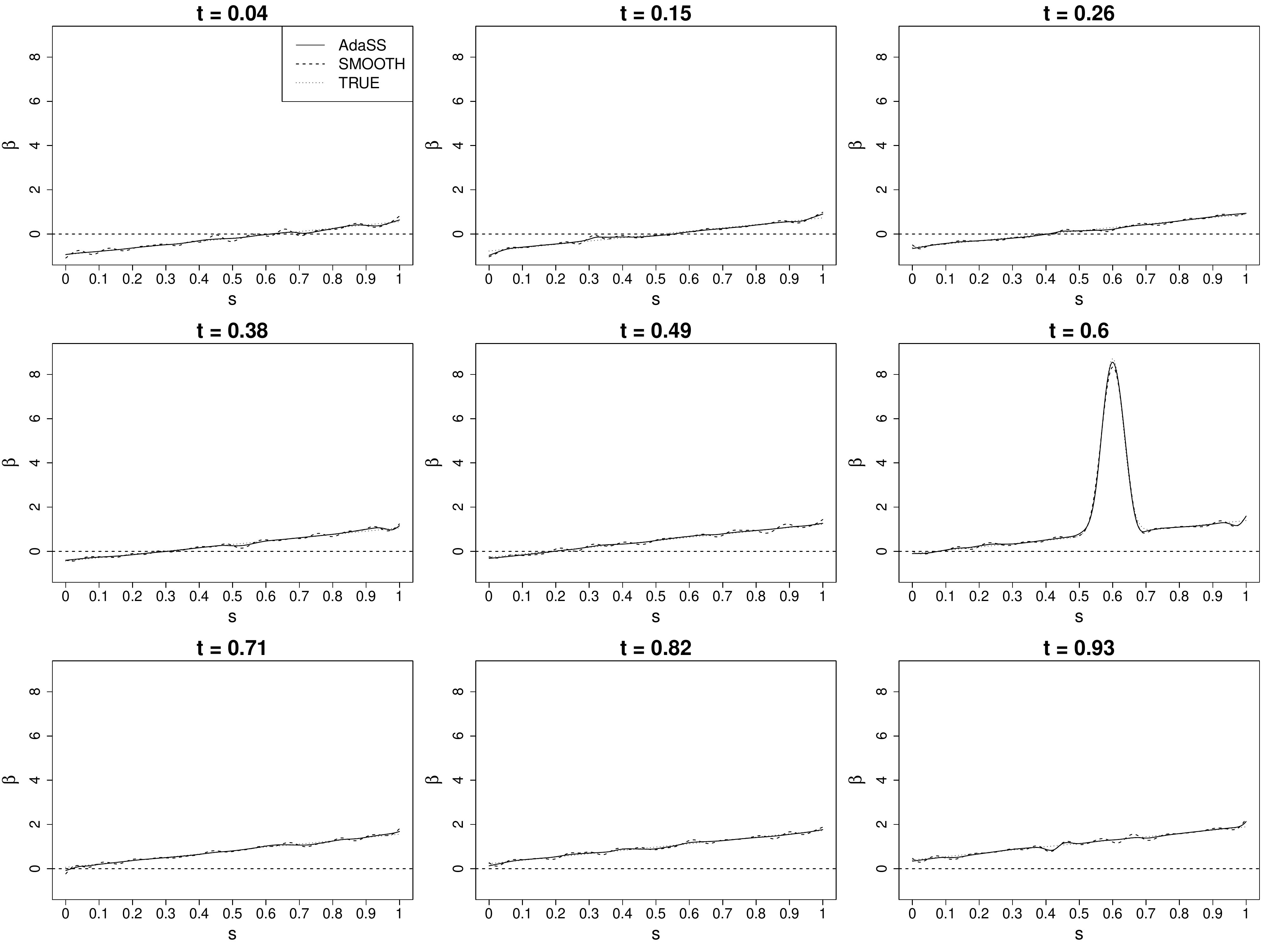}
	
	\caption{ AdaSS (solid line) and SMOOTH (dashed line) estimates of the coefficient functions and the TRUE coefficient function $\beta$ (dotted line)  for different values of $t$  in the case of the Mexican hat function. }
	\label{fig_HATest}
	
\end{figure}

\begin{figure}

	\begin{subfigure}[b]{0.49\textwidth}
		\centering
		\includegraphics[width=\textwidth]{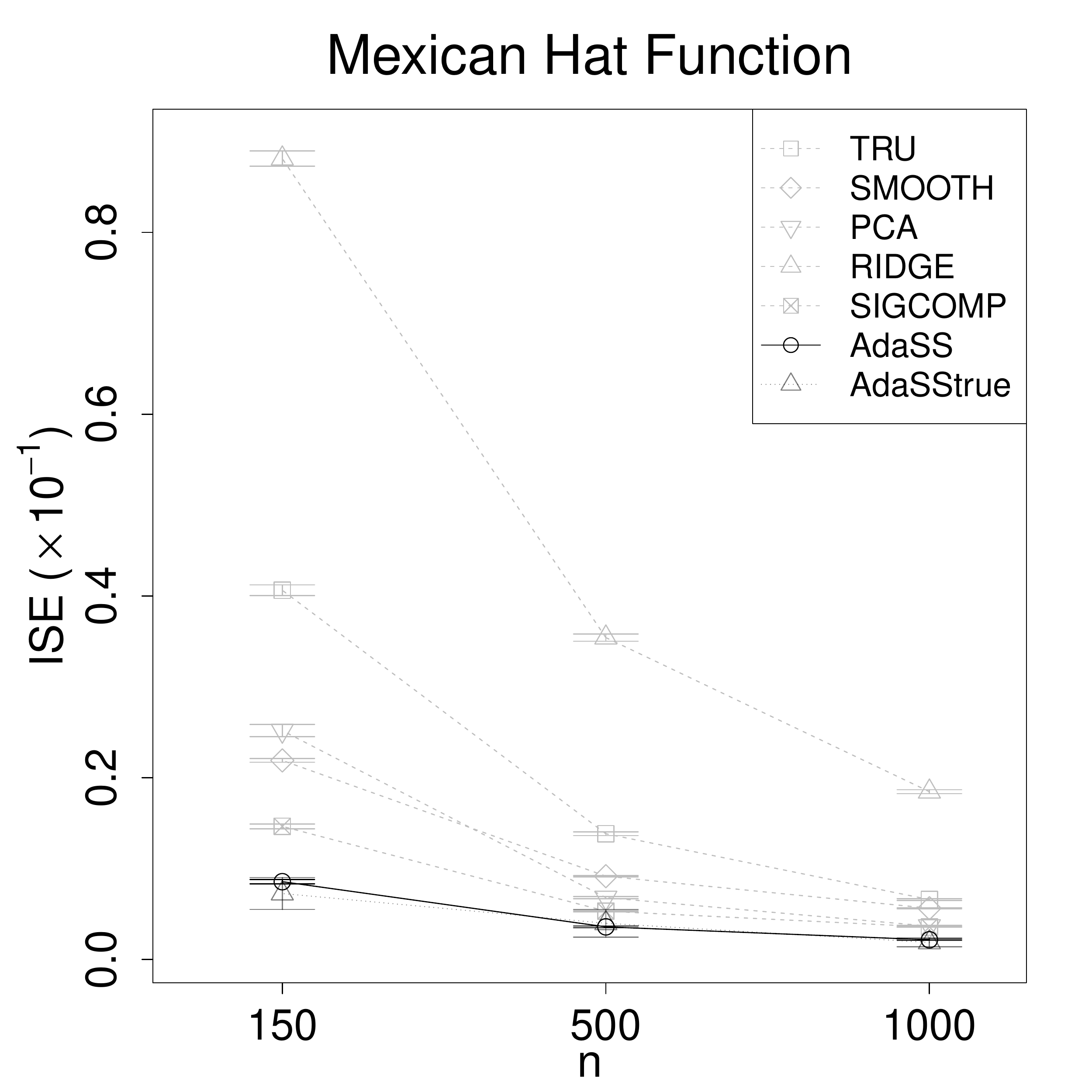}
		\caption{}
		\label{subfig_HAT_ISE}
	\end{subfigure}
	\begin{subfigure}[b]{0.49\textwidth}
		\centering
		\includegraphics[width=\textwidth]{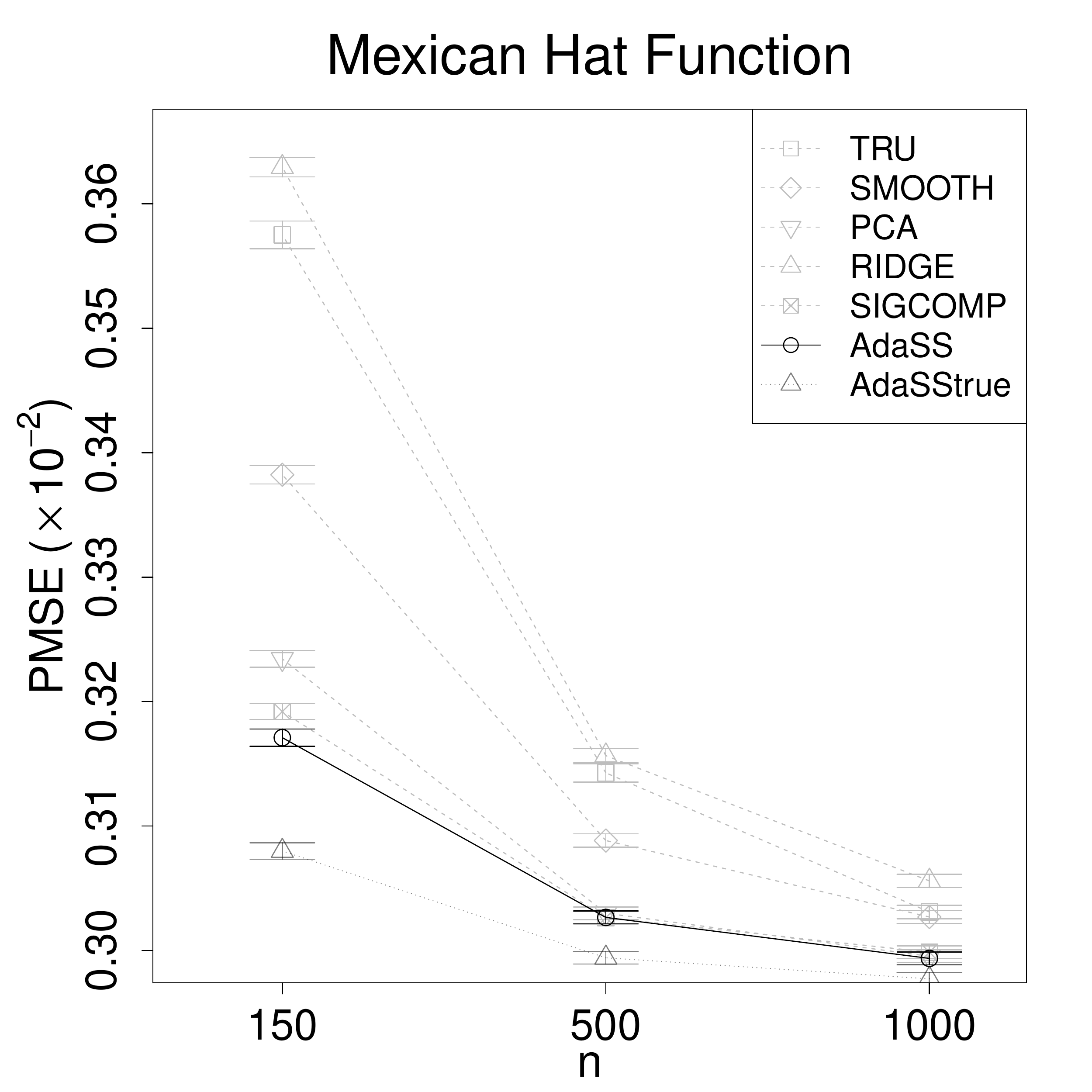}
		\caption{}
		\label{subfig_HAT_PMSE}
	\end{subfigure}

	\caption{ \subref{subfig_HAT_ISE} The integrated squared error  (ISE) and \subref{subfig_HAT_PMSE} the prediction mean squared error (PMSE)  $\pm standard\hspace{0.12cm} error$   for the TRU, SMOOTH, PCA, RIDGE, SIGCOMP, AdaSS and AdaSStrue  estimators in the case of the Mexican hat function. }
	\label{fig_HATISEPMSE}
	
\end{figure}

The Mexican hat function is a linear function with a sharp smoothness variation in central part of the domain.
In this case, the coefficient function $\beta$ is defined as 
\begin{equation*}
\beta\left(s,t\right)=-1+1.5s+1.5t+0.05\phi\left(s,t\right),\quad s,t\in\left[0,1\right]\times\left[0,1\right]
\end{equation*}
where $\phi$ is a multivariate normal distribution with mean $\bm{\mu}=\left(0.6,0.6\right)^{T}$ and diagonal covariance matrix $\bm{\Sigma}=\diag\left(0.001,0.001\right)$.
Figure \ref{fig_HATest} displays the  AdaSS and the SMOOTH estimates along with the true coefficient function for a randomly selected simulation run.
The proposed estimator tends to be smoother on the flat region and is able to better capture the peak in the  coefficient function (at $t\approx0.6$) than the SMOOTH estimate. The latter, to perform reasonably well along the whole domain, selects tuning parameters that are not sufficiently small (large) on the peaky (flat) region.
This is also  confirmed by the graphical appeal of 
 the AdaSS estimate with respect to the competitor ones.
In Figure \ref{fig_HATISEPMSE} and top of Table \ref{ta_sim}, the values of  ISE and PMSE achieved by the AdaSS, AdaSStrue, and competitor estimators are shown as functions of the sample size $n$.
Without considering the AdaSStrue estimator, the AdaSS estimator yields the lowest ISE for all sample sizes, and thus has the lowest estimation error. In terms of PMSE, it is the best one for $n=150$, whereas for $n=500,1000$ it   performs comparably with  SIGCOMP and PCA estimators.
The performance of the AdaSStrue and  AdaSS estimators is very similar in terms of ISE, whereas the  AdaSStrue shows a lower PMSE.
However, as expected, the effect of the knowledge of the true coefficient function tends to disappear as $ n $ increases, because the partial derivatives estimates become more accurate. 

\begin{table}[h]
	
	\caption{The integrated squared error (ISE) and the prediction mean squared error (PMSE) for the TRU, SMOOTH, PCA, RIDGE, SIGCOMP, AdaSS and AdaSStrue estimators. The numbers outside the parentheses are the averages  over $100$ Monte Carlo replications, and the numbers inside parentheses are  the corresponding standard errors.
		The values corresponding to the AdaSStrue estimator are emphasized to underline the fact that they rely on the knowledge of the true coefficient function, which is unlikely in real applications. In bold are marked the lowest values among the AdaSS and the competitors. }
	\label{ta_sim}
	\resizebox{\textwidth}{!}{
		\begin{tabular}{lccccccc}
			\toprule
			
			&\multicolumn{2}{c}{$n=100$} &\multicolumn{2}{c}{$n=500$}&\multicolumn{2}{c}{$n=1000$}\\
			\midrule
			& ISE ($\times 10^{-1}$) & PMSE ($\times 10^{-2}$) & ISE ($\times 10^{-1}$)& PMSE ($\times 10^{-2}$) & ISE ($\times 10^{-1}$)&PMSE ($\times 10^{-2}$) \\ 
			\midrule
			&\multicolumn{6}{c}{Mexican hat}\\
			TRU&0.4063(0.0059)&0.3575(0.0011)&0.1384(0.0020)&0.3143(0.0007)&0.0660(0.0011)&0.3031 (0.0005)\\ 
			SMOOTH&0.2191(0.0020)&0.3382(0.0007)&0.0917(0.0008)&0.3088(0.0005)&0.0564(0.0006)&0.3027 (0.0005)\\ 
			PCA&0.2519(0.0068)&0.3234(0.0007)&0.0681(0.0013)&0.3030(0.0005)&0.0368(0.0008)&0.2995 (0.0005)\\ 
			RIDGE&0.8813(0.0083)&0.3629(0.0008)&0.3542(0.0041)&0.3157(0.0006)&0.1847(0.0022)&0.3056 (0.0005)\\ 
			SIGCOMP&0.1465(0.0026)&0.3192(0.0006)&0.0532(0.0006)&\textbf{0.3026}(0.0005)&0.0358(0.0004)&0.2999 (0.0005)\\ 
			AdaSS&\textbf{0.0856}(0.0023)&\textbf{0.3171}(0.0007)&\textbf{0.0359}(0.0010)&0.3027(0.0005)& \textbf{0.0217}(0.0007)&\textbf{0.2994} (0.0005)\\ 
			\textit{AdaSStrue}&\textit{0.0726(0.0176)}&\textit{0.3080(0.0007)}&\textit{0.0399(0.0153)}&\textit{0.2994(0.0005)}&\textit{0.0188(0.0048)}&\textit{0.2977 (0.0005)}\\     [0.2cm ]
			&\multicolumn{6}{c}{Dampened harmonic}\\
			TRU&0.2851 (0.0050)&0.5403 (0.0014)&0.0983 (0.0010)&0.5051 (0.0010)&0.0651 (0.0009)&0.4960 (0.0010)\\ 
			SMOOTH&0.2288 (0.0042)&0.5391 (0.0013)&0.0836 (0.0007)&0.5032 (0.0010)&0.0555 (0.0005)&0.4936 (0.0010)\\ 
			PCA&0.3710 (0.0093)&0.5259 (0.0012)&0.1100 (0.0020)&\textbf{0.4994} (0.0010)&0.0594 (0.0011)&\textbf{0.4915} (0.0010)\\ 
			RIDGE&1.4221 (0.0135)&0.5925 (0.0016)&0.6082 (0.0076)&0.5203 (0.0011)&0.3271 (0.0038)&0.5014 (0.0010)\\ 
			SIGCOMP&0.2541 (0.0045)&\textbf{0.5221} (0.0012)&0.1235 (0.0013)&0.5018 (0.0010)&0.0942 (0.0009)&0.4950 (0.0010)\\ 
			AdaSS&\textbf{0.1749} (0.0038)&0.5241 (0.0012)&\textbf{0.0695} (0.0012)&0.4997 (0.0010)&\textbf{0.0461} (0.0008)&0.4918 (0.0010)\\ 
			\textit{AdaSStrue}&\textit{0.1504 (0.0030)}&\textit{0.5179 (0.0012)}&\textit{0.0744 (0.0018)}&\textit{0.4985 (0.0010)}&\textit{0.0582 (0.0022)}&\textit{0.4912 (0.0010)}\\  [0.2cm ]
			&\multicolumn{6}{c}{Rapid change}\\
			TRU&1.9910(0.0278)&4.0461(0.0001)&0.9178(0.0100)&3.7583(0.0001)&0.6020(0.0074)&3.6989 (0.0001)\\ 
			SMOOTH&1.2961(0.0133)&3.9427(0.0001)&0.5738(0.0046)&3.7205(0.0001)&0.3590(0.0027)&3.6787 (0.0001)\\ 
			PCA&5.1052(0.0971)&4.3070(0.0001)&1.5870(0.0271)&3.7978(0.0001)&0.8383(0.0125)&3.7141 (0.0001)\\ 
			RIDGE&10.4781(0.1059)&4.4295(0.0001)&4.1991(0.0537)&3.8459(0.0001)&2.2250(0.0278)&3.7356 (0.0001)\\ 
			SIGCOMP&1.7129(0.0209)&4.0352(0.0001)&0.8615(0.0234)&3.7702(0.0001)&0.8552(0.0167)&3.7428 (0.0001)\\ 
			AdaSS&\textbf{1.0482}(0.0166)&\textbf{3.8737}(0.0001)&\textbf{0.4526}(0.0077)&\textbf{3.6928}(0.0001)&\textbf{0.2916}(0.0044)&\textbf{3.6662} (0.0001)\\ 
			\textit{AdaSStrue}&\textit{0.8181(0.0191)}&\textit{3.8274(0.0001)}&\textit{0.3434(0.0080)}&\textit{3.6759(0.0001)}&\textit{0.2114(0.0050)}&\textit{3.6541 (0.0001)}\\ 
			[0.2cm ]
			\bottomrule
		\end{tabular}
	}
\end{table}

\subsection{Dampened Harmonic Motion Function}
\label{sec_damp}
\begin{figure}
	
	\centering
	\includegraphics[width=1\textwidth]{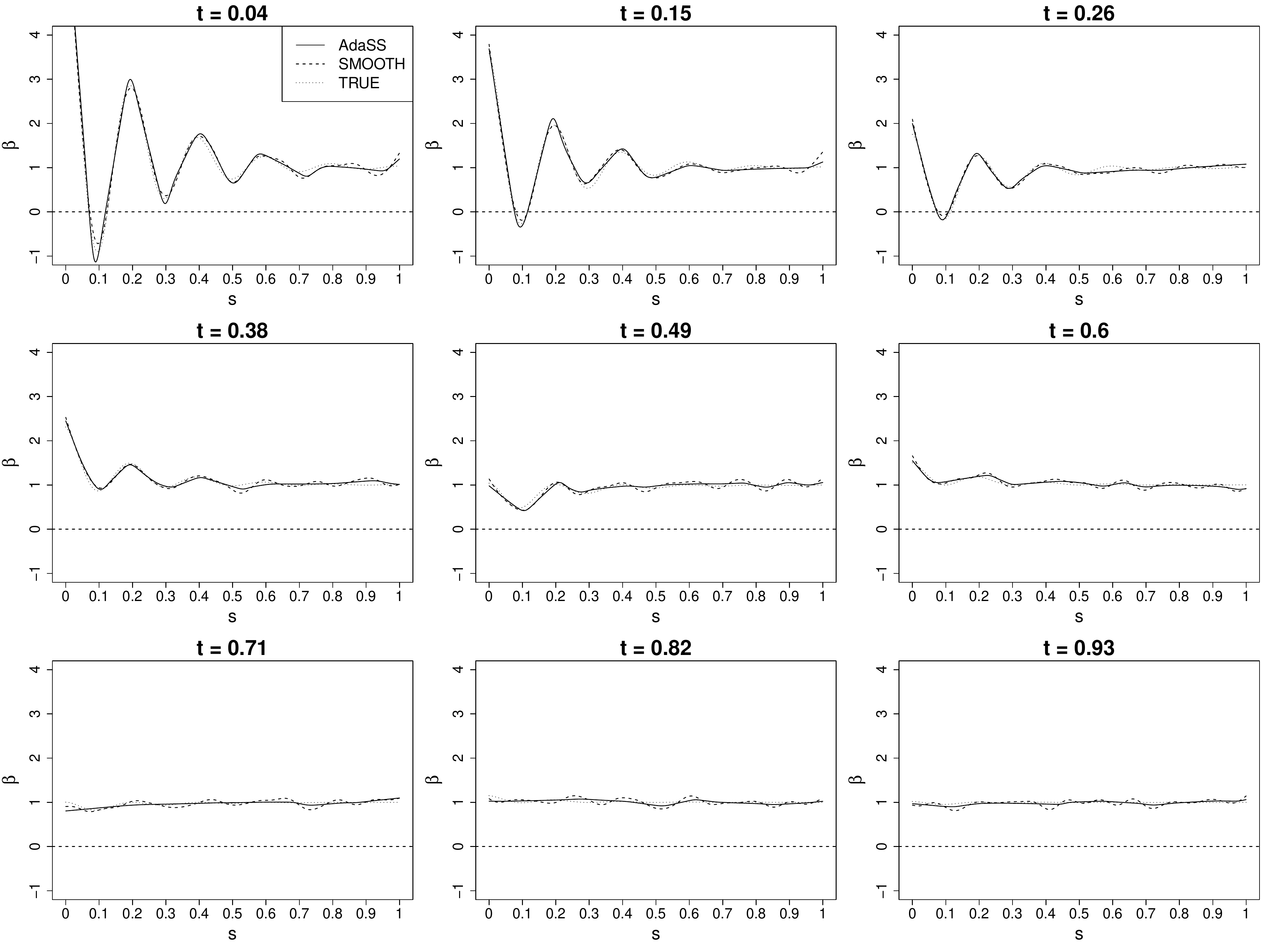}
	
	\caption{ AdaSS (solid line) and SMOOTH (dashed line) estimates of the coefficient functions and the TRUE coefficient function $\beta$ (dotted line)  for different values of $t$  in the case of the dampened harmonic motion function. }
	\label{fig_DAMPest}
	
\end{figure}

This simulation scenario considers as coefficient function $\beta$ the dampened harmonic motion function, also known as the \textit{spring function} in the engineering literature. It is characterized by a sinusoidal behaviour with exponentially decreasing amplitude, that is
\begin{equation*}
\beta\left(s,t\right)=1+5\exp\left[-5\left(s+t\right)\right]\left[\cos\left(10\pi s\right)+\cos\left(10\pi t\right)\right],\quad s,t\in\left[0,1\right]\times\left[0,1\right].
\end{equation*}
Figure \ref{fig_DAMPest} displays the AdaSS and the SMOOTH estimates along with the true coefficient function. Also in this scenario, the AdaSS estimates is smoother than the SMOOTH estimates in regions of small curvature. But, it is more flexible where the coefficient function is more wiggly. Note that intuitively, the SMOOTH estimator trades off its smoothness over the whole domain. Indeed, it over-smooths at small values of $s$ and $t$ and under-smooths elsewhere.

\begin{figure}

	\begin{subfigure}[b]{0.49\textwidth}
		\centering
		\includegraphics[width=\textwidth]{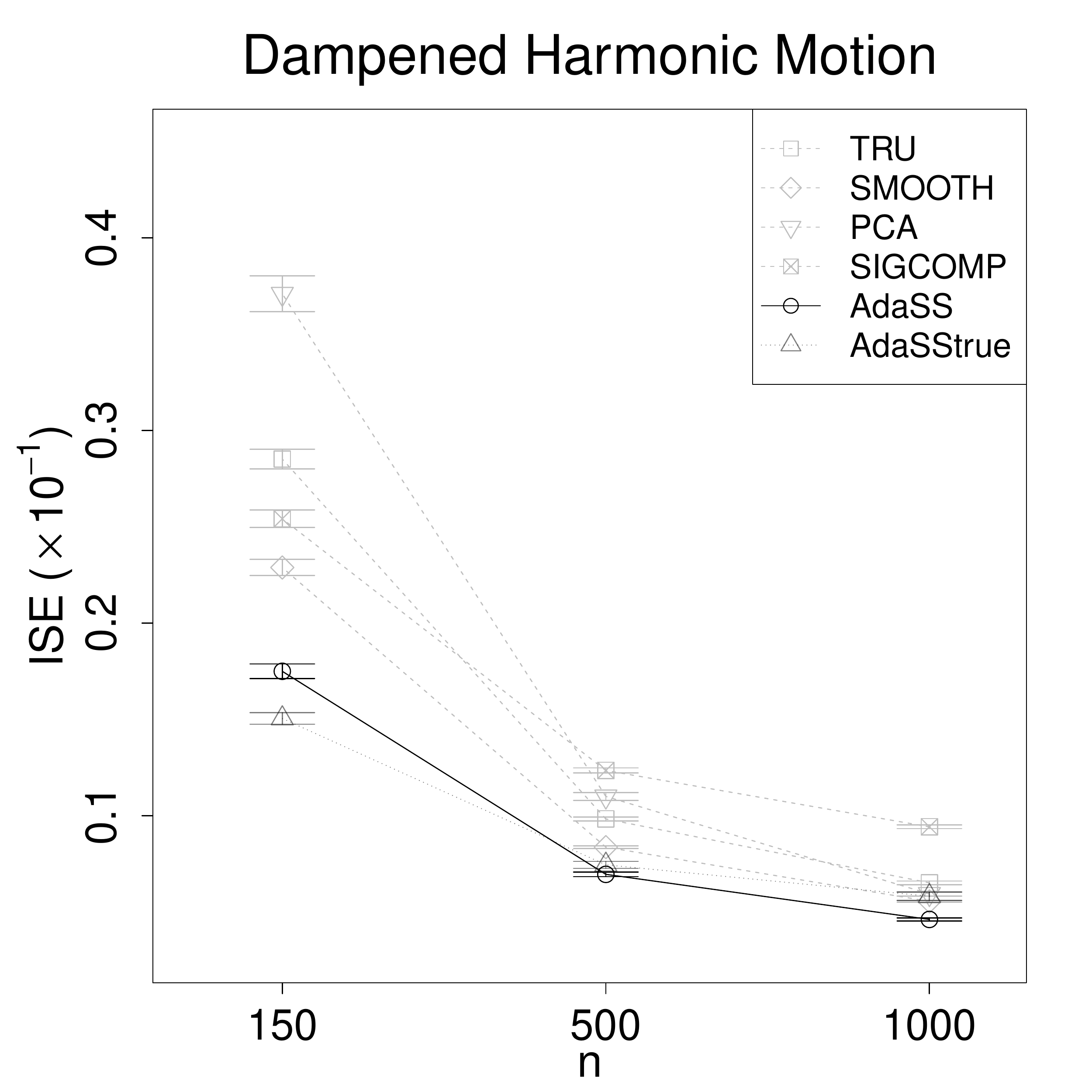}
		\caption{}
		\label{subfig_HAT_ISE}
	\end{subfigure}
	\begin{subfigure}[b]{0.49\textwidth}
		\centering
		\includegraphics[width=\textwidth]{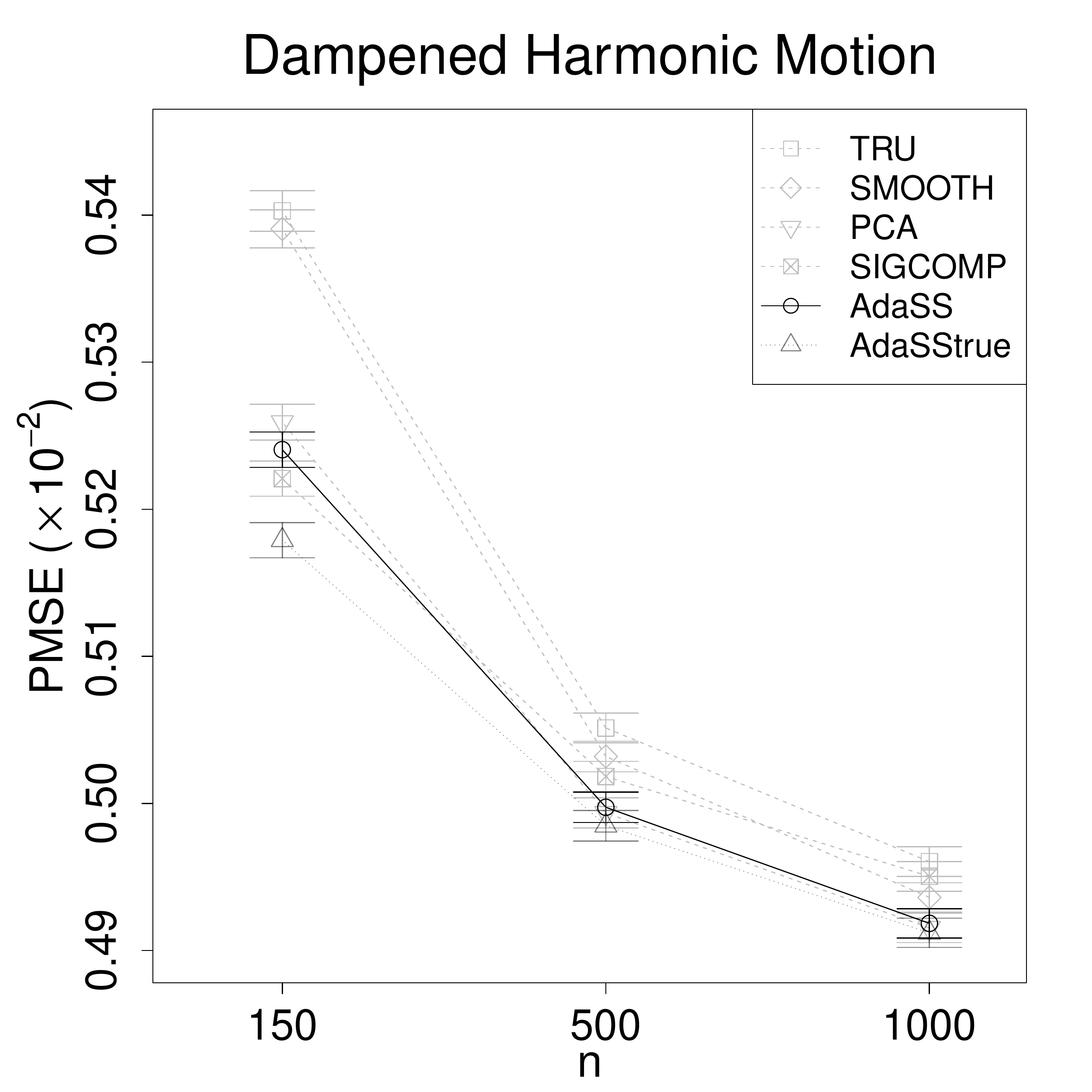}
		\caption{}
		\label{subfig_DAMP_PMSE}
	\end{subfigure}

	\caption{ \subref{subfig_HAT_ISE} The integrated squared error  (ISE) and \subref{subfig_HAT_PMSE} the prediction mean squared error (PMSE)  $\pm standard\hspace{0.12cm} error$   for the TRU, SMOOTH, PCA,  SIGCOMP, AdaSS and AdaSStrue  estimators in the case of the dampened harmonic motion function. The Ridge estimator is not considered due to its too different performance.}
	\label{fig_DAMPISEPMSE}
	
\end{figure}

In Figure \ref{fig_DAMPISEPMSE} and in the second tier of Table \ref{ta_sim}, values of the ISE and PMSE  for the AdaSS, AdaSStrue, and competitor estimators are shown as function of the sample size $n$, in the case of the dampened harmonic motion function.
Even in this case, the AdaSS estimator achives the lowest ISE for all sample sizes, and thus,  the lowest estimation error, without taking into account the AdaSStrue estimator.
Strictly speaking, in terms of PMSE, note that the proposed estimator is not always the best choice, but it shows only a small difference with best methods, viz.,  PCA and SIGCOMP estimators. 
In this case, the  AdaSS and  AdaSStrue performance is very similar for $ n=500,1000 $, whereas, for $ n=150 $, the AdaSStrue performs slightly better especially in terms of PMSE.

\subsection{Rapid Change Function}
\label{sec_rchange}
\begin{figure}
	
	\centering
	\includegraphics[width=1\textwidth]{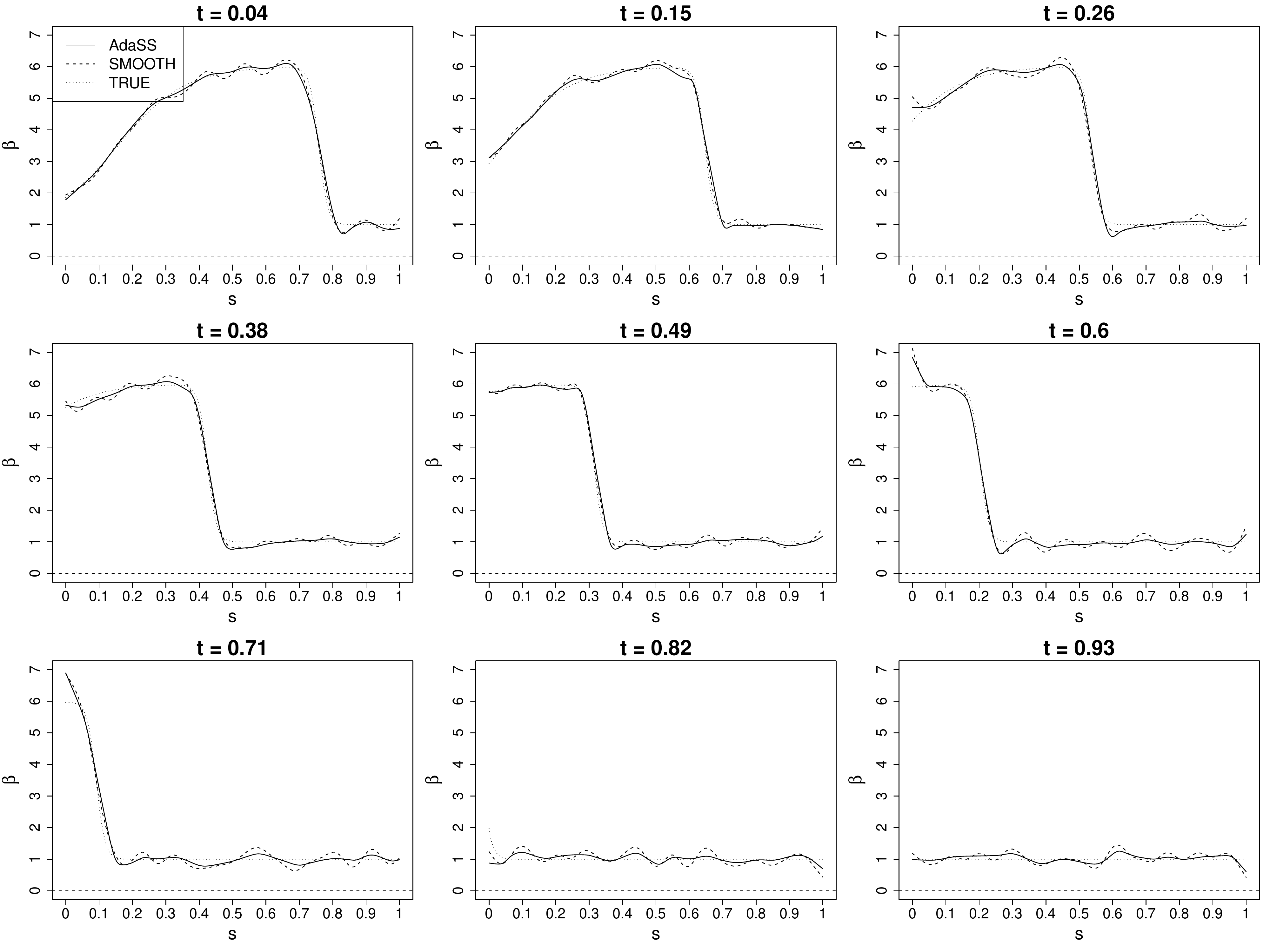}
	
	\caption{ AdaSS (solid line) and SMOOTH (dashed line) estimates of the coefficient functions and the TRUE coefficient function $\beta$ (dotted line)  for different values of $t$  in the case of the rapid change function. }
	\label{fig_RCHANGEest}
	
\end{figure}

In this scenario the true coefficient function $\beta$ is obtained by the rapid change function, that is
\begin{equation*}
\beta\left(s,t\right)=1-\frac{5}{1+\exp\left[10\left(s+t-0.2\right)\right]}+\frac{5}{1+\exp\left[75\left(s+t-0.8\right)\right]},\quad s,t\in\left[0,1\right]\times\left[0,1\right].
\end{equation*}
Figure \ref{fig_RCHANGEest} shows the AdaSS and SMOOTH estimate when $\beta$ is  the rapid change function.
The SMOOTH estimate is rougher than the AdaSS one in regions that are  far from the rapid change point. On the contrary, the AdaSS estimate is able to be smoother in the flat region and to be as accurate as the SMOOTH estimate near the rapid change point.

\begin{figure}

	\begin{subfigure}[b]{0.49\textwidth}
		\centering
		\includegraphics[width=\textwidth]{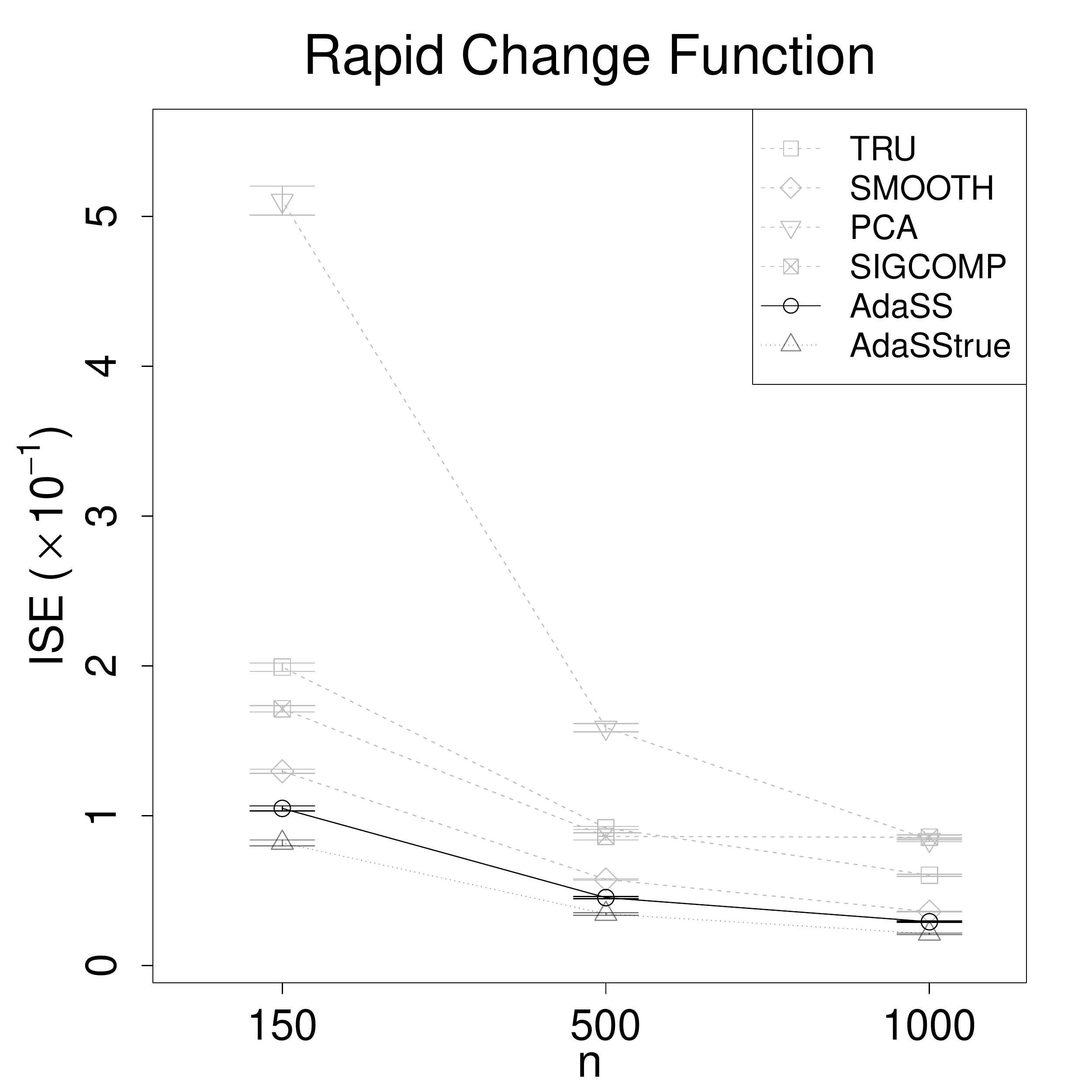}
		\caption{}
		\label{subfig_RCHANGE_ISE}
	\end{subfigure}
	\begin{subfigure}[b]{0.49\textwidth}
		\centering
		\includegraphics[width=\textwidth]{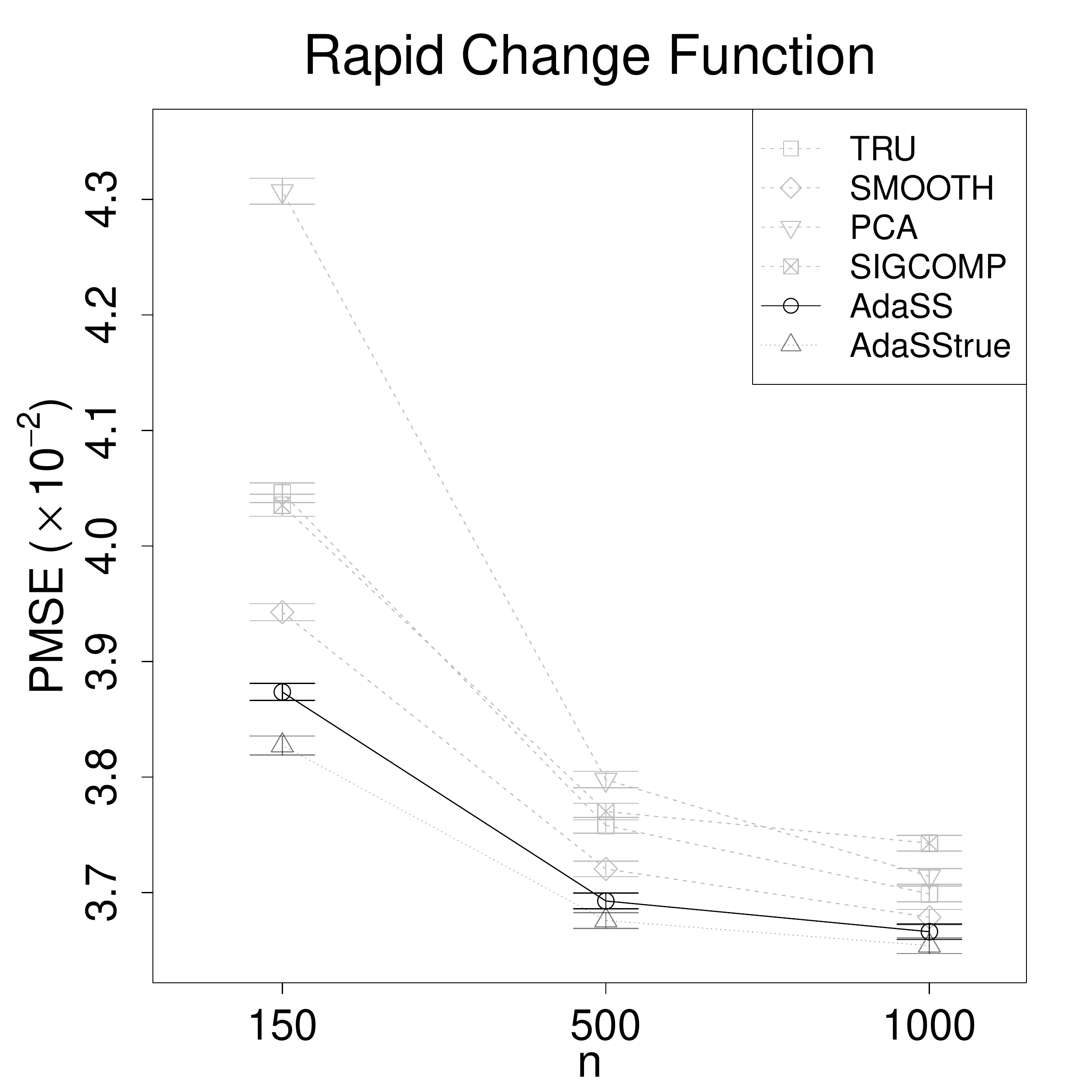}
		\caption{}
		\label{subfig_RCHANGE_PMSE}
	\end{subfigure}

	\caption{ \subref{subfig_HAT_ISE} The integrated squared error  (ISE) and \subref{subfig_HAT_PMSE} the prediction mean squared error (PMSE)  $\pm standard\hspace{0.12cm} error$   for the TRU, SMOOTH, PCA,  SIGCOMP, AdaSS and AdaSStrue  estimators in the case of the rapid change function function. The Ridge estimator is not considered due to its too different performance.}
	\label{fig_RCHANGEISEPMSE}
	
\end{figure}

In Figure \ref{fig_RCHANGEISEPMSE} and  the third tier of Table \ref{ta_sim}, values of the ISE and PMSE  for the AdaSS, AdaSStrue, and competitor estimators are shown for  sample sizes $n=150,500,1000$.
In this case, the AdaSS estimator outperforms the competitors, both in terms of ISE and PMSE.
Also in this case, the performance of the  AdaSStrue estimator is slightly better than that of the AdaSS one and this difference in performance reduces as $ n $ increases.

\section{Real-data Examples}
\label{sec_real}
In this section, two real datasets, namely \textit{Swedish mortality} and  \textit{ship CO\textsubscript{2} emission} datasets, are considered in order to asses the performance of the AdaSS estimator in real applications.

\subsection{Swedish Mortality Dataset}
\label{sec_realsw}
\begin{figure}[H]
	\centering
	\includegraphics[width=0.5\textwidth]{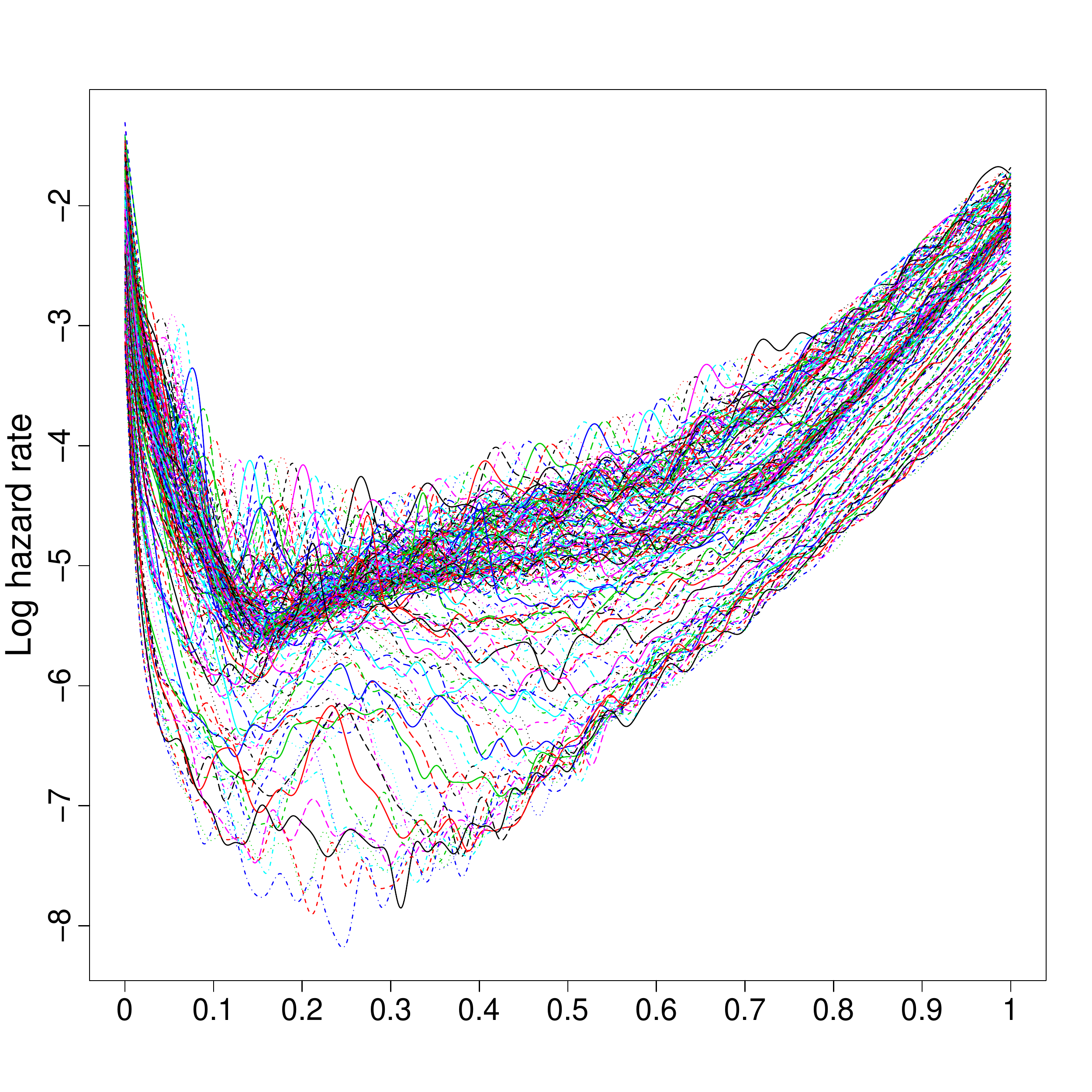}
	\caption{Log-hazard rate functions  for Swedish female cohorts from 1751 to 1935.}
	\label{fig_datasw}
\end{figure}

\begin{figure}
	
	\centering
	\includegraphics[width=1\textwidth]{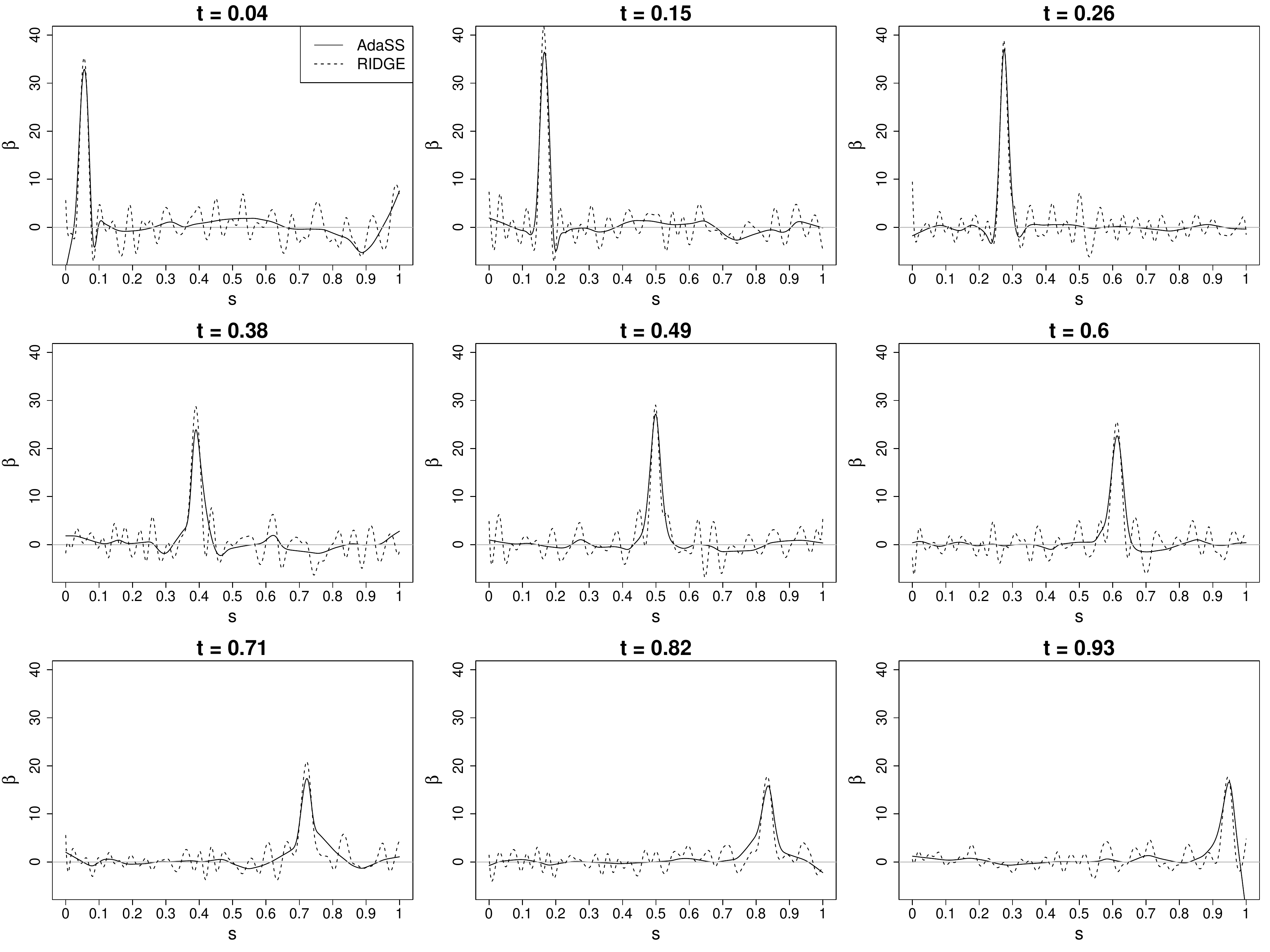}
	
	\caption{ AdaSS (solid line) and RIDGE (dashed line) estimates of the coefficient functions   for different values of $t$  in the Swedish Mortality dataset. }
	\label{fig_swest}
	
\end{figure}

The Swedish mortality dataset (available from the Human Mortality Database ---\url{http://mortality.org}---) is very well known in the functional literature as benchmark dataset. It has been analysed by Chiou and M{\"u}ller \cite{chiou2009modeling} and Ramsay
et al. \cite{ramsay2009functional}, among others.
In this analysis, we consider the log-hazard rate functions of the Swedish females mortality data for year-of-birth cohorts that refer to females born in the years 1751-1935 with ages 0-80. The value of a log-hazard rate function at a given age is the natural logarithm of the ratio of  females  died at that age and the number of females alive with the same age. The 184 considered  log-hazard rate functions \citep{chiou2009modeling} are shown in Figure \ref{fig_datasw}. Without loss of generality they have been normalized to the domain $\left[0,1\right]$.
The functions from 1751 (1752) to 1934 (1935) are considered as  observations $X_i$ ($Y_i$) of the predictor (response) in \eqref{eq_lm}, $ i=1,\dots,184 $. In this way, the  relationship between two consecutive log-hazard rate functions becomes the focus of the analysis.
To asses the predictive performance of the methods considered in the simulation study (Section \ref{sec_simulation}),  for 100 times, 166 observations out of 184 are randomly chosen, as training set, to fit the model. The 18 remaining ones are used  as test set to calculate the PMSE. 
The averages and standard deviations of  PMSEs are shown in the first line of Table \ref{ta_case}. The AdaSS estimator outperforms all the competitors. Only the RIDGE estimator has comparable predictive performance.
\begin{table}[h]
	
	\caption{The prediction mean squared error (PMSE) for the TRU, SMOOTH, PCA, RIDGE, SIGCOMP, and AdaSS
		estimators. The numbers outside the parentheses are the averages of the PMSE over $100$  replications, and the numbers inside parentheses are  the corresponding standard errors. }
	\label{ta_case}
	\resizebox{\textwidth}{!}{
		\begin{tabular}{lccccccc}
			\toprule
			
			& TRU     &   SMOOTH         &  PCA     &    RIDGE    &   SIGCOMP &AdaSS\\
			\midrule
			Swedish mortality ($\times10^{-2}$) &0.7373 (0.0000) &0.5938 (0.0000) &0.6131 (0.0000) & 0.5749  (0.0000)& 1.0173 (0.0000)& \textbf{0.5706} (0.0000)\\
			[0.2cm ]
			Ship CO\textsubscript{2} emission &0.1019 (0.0008) &    0.0814 (0.0007) &  0.0689 (0.0008)&   \textbf{0.0625}   (0.0007)& 0.1033 (0.0013)&    0.0771 (0.0007)\\
			\bottomrule
		\end{tabular}
	}
\end{table}
Figure \ref{fig_swest} shows the AdaSS estimates along with the RIDGE estimates that represents the    best competitor methods in terms of PMSE.
The proposed estimator has slightly better performance than the competitor, but, at the same time, it is much more interpretable. In fact, it is much  smoother where the coefficient function seem to be mostly flat and successfully captures the pattern of $\beta$ in the peak region. On the contrary, the RIDGE estimates is particularly rough over region of low curvature.

\subsection{Ship CO\textsubscript{2} Emission Dataset}
\label{sec_realgr}
\begin{figure}[H]
	
	\centering
	\includegraphics[width=0.4\textwidth]{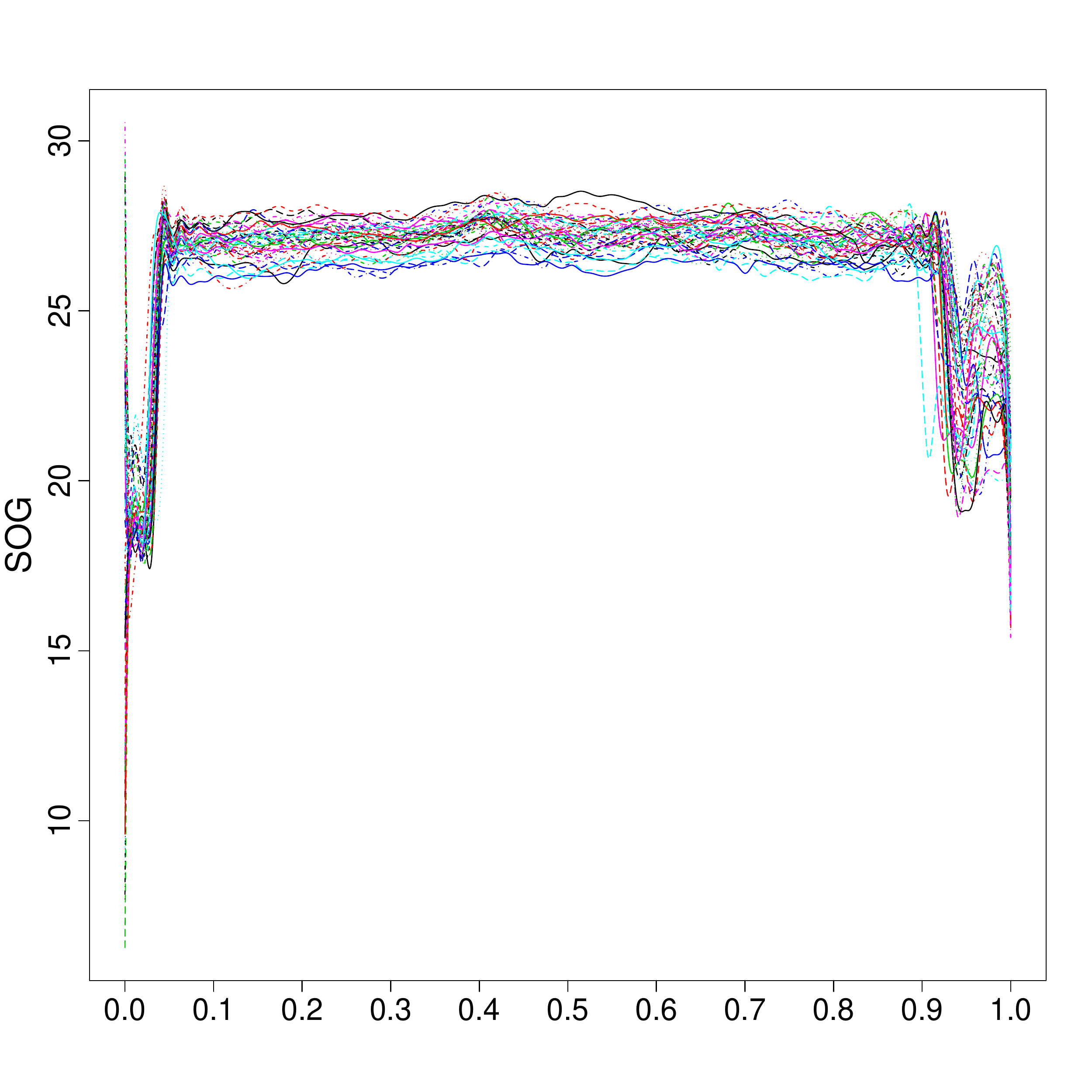}
	\includegraphics[width=0.4\textwidth]{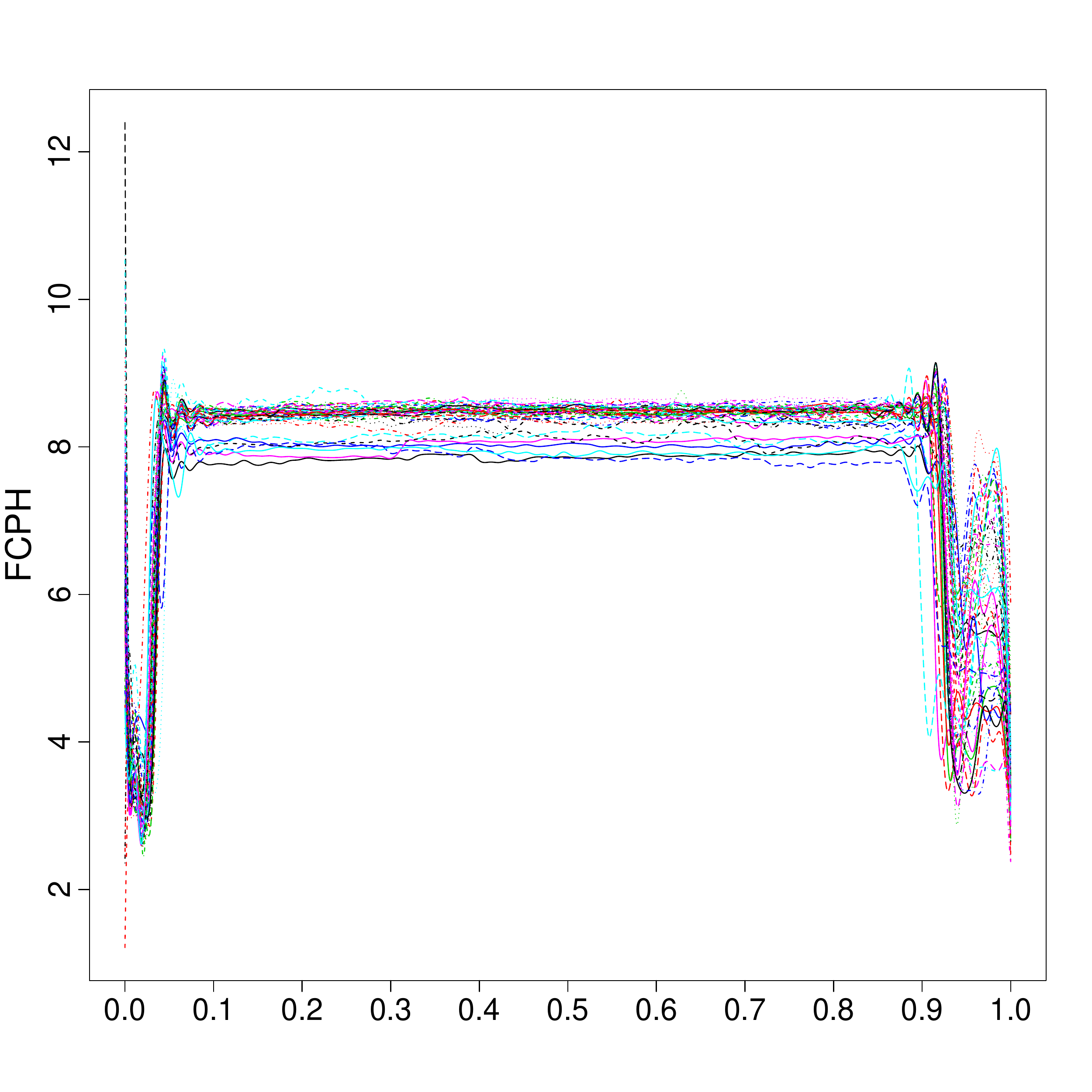}

	\caption{SOG and FCPH observations from a Ro-Pax ship.}
	\label{fig_datagrimald}
	
\end{figure}

\begin{figure}
	
	\centering
	\includegraphics[width=1\textwidth]{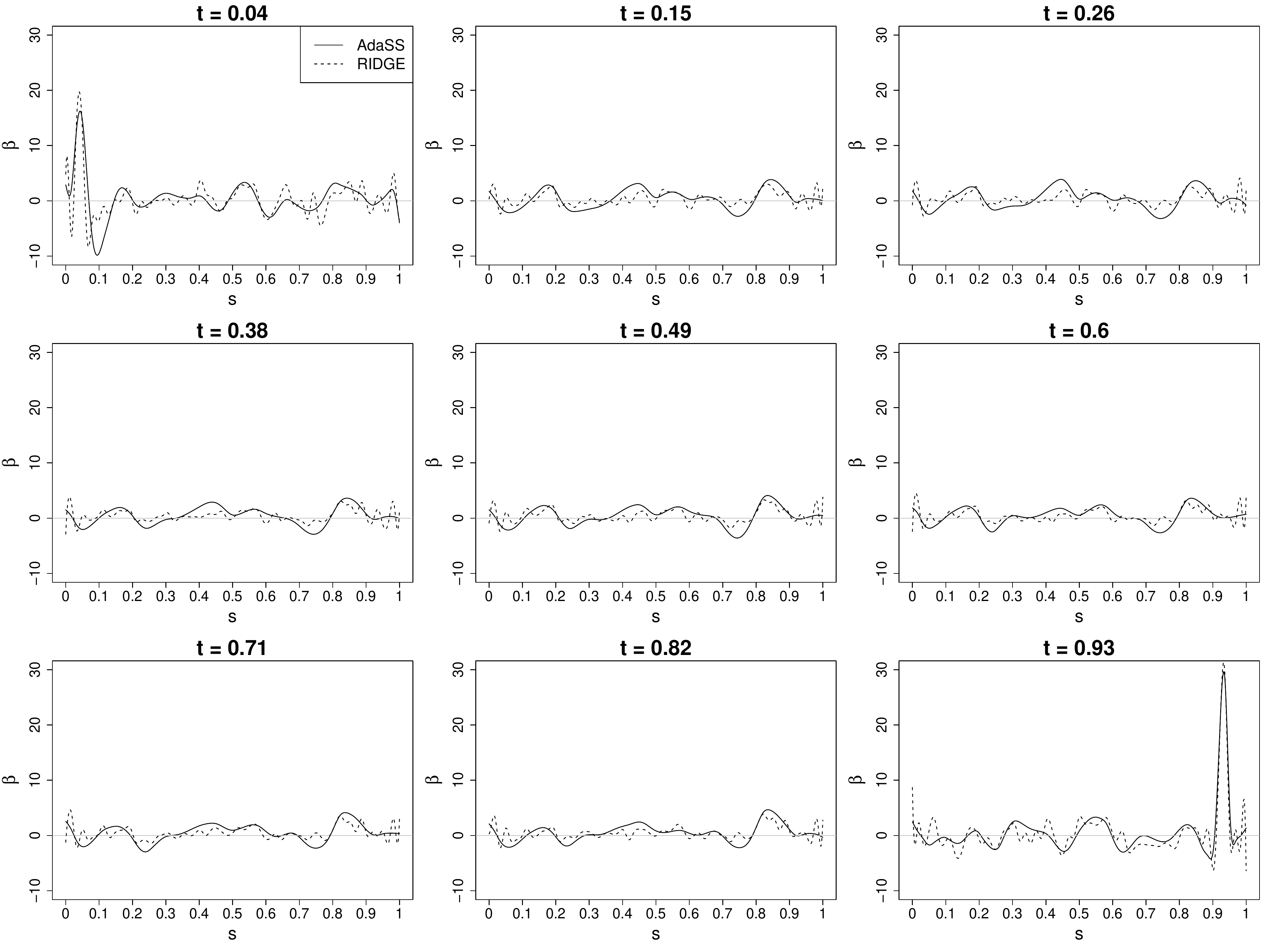}
	
	\caption{ AdaSS (solid line) and RIDGE (dashed line) estimates of the coefficient functions  for different values of $t$  in the ship CO\textsubscript{2} emission dataset. }
	\label{fig_grest}
	
\end{figure}

The ship CO\textsubscript{2} emission dataset has been thoroughly studied in the very last years \citep{lepore2018analysis,reis2019predicting,capezza2020control,centofanti2020functional}. It was provided by  the shipping company Grimaldi Group to address some aspects that are related to the issue of monitoring fuel consumptions or CO\textsubscript{2} emissions for Ro-Pax ship that sail along a route in the Mediterranean Sea.
In particular, we focus on the study of the relation between the \textit{fuel consumption per hour} (FCPH), assumed as the response,  and the \textit{speed over ground} (SOG), assumed as predictor. The observations considered were recorded from 2015 to 2017.
Figure \ref{fig_datagrimald} shows the 44 available observations of SOG and FCPH \citep{centofanti2020functional}.
Similarly to  the Swedish mortality dataset, The prediction performance of the methods are assessed   by randomly chosen 40 out of 44 observations  to fit the model and by using the 4 remaining observations to compute the PMSE. This is repeated 100 times. 
The averages and standard deviations of the PMSEs are listed in the second line of Table \ref{ta_case}.
The AdaSS estimator is  in this case outperformed by the RIDGE estimator, which achieves the lowest PMSE.
However, as shown in  Figure \ref{fig_grest},  it is able both to well estimate the coefficient function over peaky regions, as the RIDGE estimator, and to smoothly adapt over the remaining  part of the domain.
In this case, also the PCA estimator achieves smaller   PMSE than that of the proposed estimator. However, the PCA estimator is even rougher than the RIDGE estimator and, thus, it is not shown in  Figure \ref{fig_grest}.

\section{Conclusion}
\label{sec_conclusion}
In this article, the AdaSS estimator is proposed for the function-on-function linear regression model where  each value of the response, for any domain point, depends linearly on the full trajectory of the predictor. The introduction of two adaptive smoothing penalties, based on initial estimate of its partial derivatives,  allows the proposed estimator to better adapt to  the coefficient function. 
By means of a simulation study, the proposed estimator has proven favourable performance with respect to those achieved by the five competitors already appeared in the literature before, both in terms of estimation and prediction error.
The adaptive  feature of the AdaSS estimator is advantageous for the interpretability of the results with respect to the competitors.
Moreover, its performance has shown to be competitive also with respect to the case where the true coefficient function is known.
Finally, the proposed estimator has been successfully applied to real-data examples, viz., the Swedish mortality and ship CO\textsubscript{2} emission datasets.
However, some challenges are still open. Even though the proposed evolutionary algorithm has shown to perform particularly well both in the simulation study and the real-data examples, the choice of the tuning parameters still remains in fact a critical issue, because of the curse of  dimensionality.
This could be even more problematic in the perspective to extend the AdaSS estimator to the FoF regression model with multiple predictors.

\appendix
\section{Approximation of the Two Penalty Terms for the AdaSS Estimator Derivation }
\label{sec_app1}
In this section the approximations of \eqref{eq_pen1} and \eqref{eq_pen2} are obtained.
For the first penalty, by using \eqref{eq_app1}, we have
\begin{multline}
\lambda^{AdaSS}_{s}\int_{\mathcal{S}}\int_{\mathcal{T}}\frac{1}{\left(|\widehat{\beta_{s}^{m_{s}}}\left(s,t\right)|+\delta_s\right)^{\gamma_s}}\left(\mathcal{L}_{s}^{m_{s}}\alpha\left(s,t\right)\right)^{2}dsdt\\
\approx \lambda^{AdaSS}_{s}\int_{\mathcal{S}}\int_{\mathcal{T}}\frac{1}{\left(|\sum_{i=1}^{L_s+1}\sum_{j=1}^{L_t+1}\widehat{\beta_{s}^{m_{s}}}\left(\tau_{s,i},\tau_{t,j}\right)I_{\left[\left(\tau_{s,i-1},\tau_{s,i}\right)\times \left(\tau_{t,j-1},\tau_{t,j}\right)\right]}\left(s,t\right)|+\delta_s\right)^{\gamma_s}}\left(\mathcal{L}_{s}^{m_{s}}\alpha\left(s,t\right)\right)^{2}dsdt\\
=\lambda^{AdaSS}_{s}\sum_{i=1}^{L_s+1}\sum_{j=1}^{L_t+1}\int_{\left[\tau_{s,i-1},\tau_{s,i}\right]}\int_{\left[\tau_{t,j-1},\tau_{t,j}\right]}\frac{1}{\left(|\widehat{\beta_{s}^{m_{s}}}\left(\tau_{s,i},\tau_{t,j}\right)|+\delta_s\right)^{\gamma_s}}\left(\mathcal{L}_{s}^{m_{s}}\alpha\left(s,t\right)\right)^{2}dsdt\\
=\lambda^{AdaSS}_{s}\sum_{i=1}^{L_s+1}\sum_{j=1}^{L_t+1}d^{s}_{ij}\int_{\left[\tau_{s,i-1},\tau_{s,i}\right]}\int_{\left[\tau_{t,j-1},\tau_{t,j}\right]}\left(\mathcal{L}_{s}^{m_{s}}\alpha\left(s,t\right)\right)^{2}dsdt,
\end{multline}
where $d^{s}_{ij}=\Big\{\frac{1}{\left(|\widehat{\beta_{s}^{m_{s}}}\left(\tau_{s,i},\tau_{t,j}\right)|+\delta_s\right)^{\gamma_s}}\Big\}$.
Then, for \eqref{eq_betaapp0}, and following Ramsay and
Silverman  \cite{ramsay2005functional},  pag. 292,
\begin{multline}
\lambda^{AdaSS}_{s}\sum_{i=1}^{L_s+1}\sum_{j=1}^{L_t+1}d^{s}_{ij}\int_{\left[\tau_{s,i-1},\tau_{s,i}\right]}\int_{\left[\tau_{t,j-1},\tau_{t,j}\right]}\left(\mathcal{L}_{s}^{m_{s}}\alpha\left(s,t\right)\right)^{2}dsdt\\
= \lambda^{AdaSS}_{t}\sum_{i=1}^{L_s+1}\sum_{j=1}^{L_t+1}d^{t}_{ij}\Tr\left[ \bm{B}_{\alpha}^{T}\bm{W}_{s,i}\bm{B}_{\alpha}\bm{R}_{t,j}\right]
\end{multline}
where  $\bm{W}_{s,i}=\int_{\left[\tau_{s,i-1},\tau_{s,i}\right]}\bm{\psi}^{s}\left(s\right)\bm{\psi}^{s}\left(s\right)^{T}ds$,    $\bm{R}_{t,j}=\int_{\left[\tau_{t,j-1},\tau_{t,j}\right]}\mathcal{L}_{t}^{m_{t}}\left[\bm{\psi}^{t}\left(t\right)\right]\mathcal{L}_{t}^{m_{t}}\left[\bm{\psi}^{t}\left(t\right)\right]^{T}dt$, and $d^{s}_{ij}=\Big\{\frac{1}{\left(|\widehat{\beta_{s}^{m_{s}}}\left(\tau_{s,i},\tau_{t,j}\right)|+\delta_s\right)^{\gamma_s}}\Big\}$, for $i=1,\dots,L_s+1$ and $j=1,\dots,L_t+1$.
Thus, \eqref{eq_pen1} is demonstrated, the arguments are analogous for \eqref{eq_pen2}.	



\bibliographystyle{acmtrans-ims}
\bibliography{References}
\end{document}